\documentclass[%
 preprint,
superscriptaddress,
 amsmath,amssymb,
 aps,
prl,
]{revtex4-1}

\usepackage{graphicx}
\usepackage{dcolumn}
\usepackage{bm}
\usepackage[english]{babel} 
\usepackage{hyperref}
\usepackage{physics}
\usepackage{amsmath}
\usepackage{subfigure}
\usepackage{bm}
\usepackage{epstopdf}

\begin{document}

\title{$s$-$d$ model for local and nonlocal spin dynamics in laser-excited magnetic heterostructures}

\author{M. Beens}
\email[Corresponding author: ]{m.beens@tue.nl}
\affiliation{Department of Applied Physics, Eindhoven University of Technology \\ P.O. Box 513, 5600 MB Eindhoven, The Netherlands}
\author{R.A. Duine }
\affiliation{Department of Applied Physics, Eindhoven University of Technology \\ P.O. Box 513, 5600 MB Eindhoven, The Netherlands}
\affiliation{ Institute for Theoretical Physics, Utrecht University \\
Leuvenlaan 4, 3584 CE Utrecht, The Netherlands}
\author{B. Koopmans}
\affiliation{Department of Applied Physics, Eindhoven University of Technology \\ P.O. Box 513, 5600 MB Eindhoven, The Netherlands}

\date{\today}

\begin{abstract}
We discuss a joint microscopic theory for the laser-induced magnetization dynamics and spin transport in magnetic heterostructures based on the $s$-$d$ interaction. Angular momentum transfer is mediated by scattering of itinerant $s$ electrons with the localised ($d$  electron) spins. We use the corresponding rate equations and focus on a spin one-half $d$ electron system, leading to a simplified analytical expression for the dynamics of the local magnetization that is coupled to an equation for the non-equilibrium spin accumulation of the $s$ electrons. We show that this description converges to the microscopic three-temperature model in the limit of a strong $s$-$d$ coupling. The equation for the spin accumulation is used to introduce diffusive spin transport. The presented numerical solutions show that during the laser-induced demagnetization in a ferromagnetic metal a short-lived spin accumulation is created that counteracts the demagnetization process. Moreover, the spin accumulation leads to the generation of a spin current at the interface of a ferromagnetic and non-magnetic metal. Depending on the specific magnetic system, both local spin dissipation and interfacial spin transport are able to enhance the demagnetization rate by providing relaxation channels for the spin accumulation that is built up during demagnetization in the ferromagnetic material. 
\end{abstract}

\maketitle

\section{I.$\qquad$  Introduction} 

Exciting magnetic systems with ultrashort laser pulses gives rise to  fascinating physics. First, it was shown that a femtosecond laser pulse can quench the magnetization of a ferromagnetic thin-film on a subpicosecond timescale  \cite{Beaurepaire1996}. Later, all-optical magnetization switching was discovered in GdFeCo alloys \cite{Stanciu2007}, which proved the high potential of using ultrashort laser pulses for future data writing technologies.  Moreover, it was demonstrated that the laser pulse generates a spin current \cite{malinowski2008control,Melnikov2011}. In non-collinear magnetic heterostructures the ultrafast generated spin current exerts a spin-transfer torque  \cite{Choi2014,Schellekens2014stt}, leading to the excitation of Terahertz standing spin waves \cite{Razdolski2017,lalieu2017absorption}. Understanding all these ultrafast phenomena paves the way towards faster magnetic data technologies, and bridges the boundaries between photonics, spintronics and magnonics. 

Despite the vast experimental developments within the field, the microscopic origin of the observed demagnetization rates is still heavily debated. Various microscopic processes have been proposed as being the dominant mechanism, such as  (i) the coherent interaction between the photons and the spins \cite{Zhang2000,Bigot2009}, 
 (ii) spin-dependent transport of hot electrons \cite{Battiato2010}, and (iii) local spin dynamics as triggered by laser heating or excitation \cite{Beaurepaire1996,Koopmans2005,Kazantseva2007,Krauss2009,Koopmans2010,Manchon2012,Mueller2014,Nieves2014,Tveten2015,Krieger2015,Tows2015}. In the latter case, the models often rely on the assumption that heating of the electrons increases the amount of spin-flip scattering events, resulting in the transfer of angular momentum. An example of this type of models is the microscopic three-temperature model (M3TM) \cite{Koopmans2010}, where it is assumed that the magnetization dynamics is dominated by Elliott-Yafet electron-phonon scattering. Arguably, other types of scattering mechanisms can also account for the observed demagnetizations rates, such as Elliott-Yafet electron-electron scattering \cite{Krauss2009} and electron-magnon scattering \cite{Manchon2012,Tveten2015}. The latter stems from the $s$-$d$ interaction in ferromagnetic transition metals, that couples the local magnetic moments ($d$ electrons) and free carriers ($s$ electrons). Similar models were derived to describe the ultrafast magnetization dynamics in semiconductors \cite{Cywinski2007} and ferrimagnetic alloys \cite{Gridnev2016}. 

Another important question is what mechanism drives the optically induced spin currents in magnetic heterostructures. First, it could be directly related to the proposed superdiffusive spin currents created in the magnetic material \cite{Battiato2010,Battiato2012}. Secondly, the laser-induced thermal gradients can generate a spin current resulting from the spin-dependent Seebeck effect \cite{Choi2015,Alekhin2017}. Recently, it was proposed that the spin-polarized electrons are generated at a rate given by the temporal derivative of the magnetization \cite{Choi2014}. Interestingly, this implies that the demagnetization and the generated spin current are driven by the same physical mechanism. The $s$-$d$ interaction, which mediates angular momentum transfer between the local magnetic moments and itinerant electrons, is a principal candidate \cite{Choi2014,Tveten2015,Shin2018}. 

In this work, we discuss an extended $s$-$d$ model for laser-induced magnetization dynamics that includes spin transport. The model describes that during demagnetization an out-of-equilibrium spin accumulation is created in the $s$ electron system  \cite{Tveten2015,Cywinski2007}, which leads to the generation of a spin current in magnetic heterostructures \cite{Choi2014,Shin2018}. We apply the $s$-$d$ model to investigate the interplay between the local magnetization dynamics and spin transport in laser-excited magnetic heterostructures. The numerical solutions of the rate equations show a qualitative agreement with the experiments and support the view that the $s$-$d$ interaction could be the main driving force of the observed ultrafast phenomena \cite{Tveten2015,Choi2014,Shin2018}. Furthermore, the crucial role of the spin accumulation is emphasized, namely (i) the generated spin accumulation has a negative feedback on the demagnetization process \cite{Tveten2015,Cywinski2007} and (ii) this bottleneck can be removed by either local spin-flip processes or by electron spin transport. Hence, both local and nonlocal processes play a crucial role in the magnetization dynamics. Finally, we discuss the limit in which the $s$-$d$ model becomes equivalent to the M3TM, and we conclude with an outlook. 

We start with an overview of the derivation of the $s$-$d$ model in Section II and we highlight the simplifications we use compared to the derivations reported in Ref.\ \cite{Tveten2015} and Ref.\ \cite{Cywinski2007}. Importantly, we show that the $s$-$d$ model can be written in a mathematical form analog to the M3TM. In Section III, we model the demagnetization experiments and discuss the role of the spin accumulation. We describe the laser-induced dynamics in a collinear magnetic heterostructure in Section IV. We explain how the different demagnetization rates of the parallel and antiparallel configuration can be understood from the $s$-$d$ model. In Section V,  we describe a bilayer consisting of a ferromagnetic and non-magnetic  metallic layer. Here, we introduce diffusive spin transport, similar to the modeling as presented in \cite{Choi2014,Kimling2017,Shin2018}. We specifically address the interplay between the local magnetization dynamics and spin transport. We investigate the role of the layer thickness on the magnetization dynamics and we analyze the temporal profile of the injected spin current in the non-magnetic layer.  

\section{II.$\qquad$  Model   } 

In this section, we give an overview of the derivation of the $s$-$d$ model for ultrafast magnetization dynamics in transition metal ferromagnets. Although our approach is closely related to the derivation as presented in Ref. \cite{Tveten2015}, it mathematically resembles the results for magnetic semiconductors \cite{Cywinski2007}. We keep our notation consistent with these references and we highlight the modifications that are needed to reach the simplified $s$-$d$ model that is used in the remainder of this paper. 

\begin{figure}[ht!]
\includegraphics[scale=0.95]{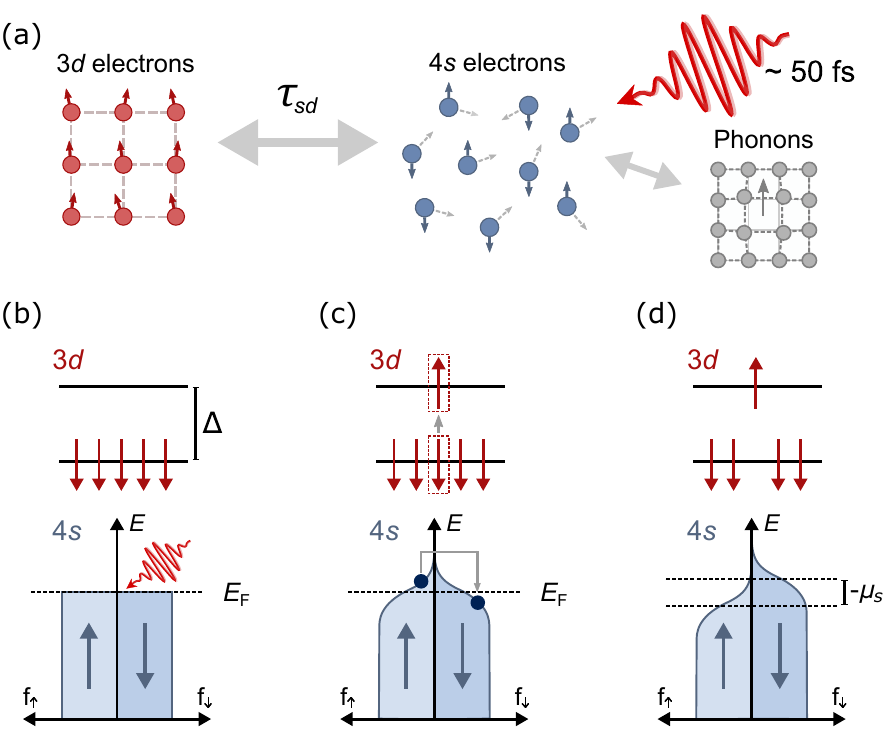}
\caption{\label{fig:fig1} Schematic overview of the $s$-$d$ model for ultrafast demagnetization \cite{Tveten2015,Cywinski2007}. (a) The system is divided into a subsystem of localised $3d$ electrons and itinerant $4s$ electrons. The laser pulse heats up the $s$ electrons. Angular momentum is transferred between the $s$ and $d$ subsystems by the $s$-$d$ interaction. Secondly, angular momentum can dissipate out of the combined system by additional spin-flip processes in the $s$ system, e.g., Elliott-Yafet electron-phonon scattering. (b)-(d) Schematically show the occupation of the energy levels in the $d$  and $s$ subsystems during the laser heating (for $S=1/2$). (b) Indicates the ground state ($T_e=0\mbox{ K}$). (c) Shows that the broadening of the Fermi-Dirac distribution allows spin-flip transitions around the Fermi level. This process is accompanied by a spin-flip of a local $d$ spin. The $s$ electrons thermalize rapidly and a non-zero spin accumulation $\mu_s$ is created, as is indicated in Fig. (d).
} 
\end{figure}

Analogous to Ref.\ \cite{Tveten2015}, we define the ferromagnetic transition metal in terms of two separate electronic systems, corresponding to the $3d$ and $4s$ electrons. A schematic overview of the model is presented in Fig. \ref{fig:fig1}(a). The $d$ electrons are the main contributor to the magnetic properties of the system and are relatively localised. Therefore, we approximate the $d$ electron system as a lattice of localised spins. At each lattice site there is only one spin and the atomic magnetic moment is given by $\mu_{\mathrm{at}}= 2 S \mu_{\mathrm{B}}$, where $\mu_{\mathrm{B}}$ is the Bohr magneton and $S$ is the spin quantum number. We neglect the orbital angular momentum. 

In this work, we describe the localised spin system within a Weiss mean field approach, similar to the description used in the M3TM \cite{Koopmans2010}. The Hamiltonian of the $d$ electrons is expressed as

\begin{eqnarray}
\label{eq:eq1} 
\hat{H}_{d} &=& \Delta  \sum_j \hat{S}^{d,z}_j ,
\end{eqnarray}

\noindent where $\hat{S}^{d,z}_j$ is the $z$ component of the spin at lattice site $j$ and $\Delta$ is the exchange splitting. Hence, each spin corresponds to a system of $2S+1$ energy levels splitted by energy $\Delta$. Note that this description of the $d$ electron system does not consider spin-wave excitations, which makes it different from the approach in Ref.\ \cite{Tveten2015}.

The $s$ electrons are described as a free electron gas. They are coupled to the localised spins through the on-site $s$-$d$ interaction, given by \cite{Tveten2015}

\begin{eqnarray}
\label{eq:eq2} 
\hat{H}_{sd} &=& J_{sd} V_{\mathrm{at}}  \sum_j \hat{\mathbf{S}}_j^d \cdot \hat{\mathbf{s}} (\mathbf{r}_j)  .  \end{eqnarray}
 
\noindent Here, $J_{sd} $ is the $s$-$d$ exchange coupling constant, $V_{\mathrm{at}}$ is the atomic volume, $\hat{\mathbf{S}}^{d}_j$ is the spin operator of the spin at lattice site $j$, and $\hat{\mathbf{s}}(\mathbf{r}_j)$ is the spin density operator of the $s$ electrons at position $\mathbf{r}_j$ of lattice site $j$.

We express $\hat{\mathbf{s}}(\mathbf{r}_j)$ in terms of the electron creation and annihilation operators in momentum space. This yields \cite{Cywinski2007}

\begin{eqnarray}
\label{eq:eq3}
\hat{H}_{sd}  &=& \sum_j \sum_{\mathbf{k}\mathbf{k}'} \Big[
 J_{j\mathbf{k}\mathbf{k}'}^*
c^\dagger_{\mathbf{k}'\downarrow} c_{\mathbf{k}\uparrow} \hat{S}^{d+}_j
+
\mbox{H.c.}
 \Big] ,
\end{eqnarray}

\noindent where the coupling strength is parametrized by the matrix element $J_{j\mathbf{k}\mathbf{k}'}$. $\hat{S}^{d\pm}_j$ corresponds to the spin ladder operator for the spin at lattice site $j$. The operator $c^\dagger_{\mathbf{k}\sigma}$ ($c_{\mathbf{k}\sigma}$) creates (annihilates) an $s$ electron with momentum $\mathbf{k}$ and spin $\sigma$. In the transition from Eq. (\ref{eq:eq2}) to Eq. (\ref{eq:eq3}) the terms proportional to the $z$ components are omitted and rewritten in terms of a mean-field energy shift in the Hamiltonian for the $s$ electrons \cite{Tveten2015}. The similar energy shift in the $d$ electron system (a shift of $\Delta$) plays a minor role and is neglected. 

Equation (\ref{eq:eq3}) describes the spin-flip scattering of $s$ electrons with the localised spins, which mediates angular momentum transfer between the $s$ and $d$ electron systems, but conserves the total angular momentum. Hence, these scattering events change the total spin in the $z$ direction of the $d$ electron system. To calculate the resulting magnetization dynamics, we apply perturbation theory using the density matrix formalism. We only show the most important steps, for more details we refer to Ref.\ \cite{Cywinski2007}, where an equivalent calculation is presented for semiconductors. In contrast to Ref.\ \cite{Cywinski2007}, our system does include a direct ($d$-$d$) exchange interaction between the localised spins, as represented by Eq. (\ref{eq:eq1}). 

First, we assume that the density matrix of the complete system can be factorized in terms of a density matrix $\hat{\rho}^C$ for the carriers ($s$ electrons) and $\hat{\rho}^S$ for the localised spins ($d$ electrons). Secondly, we assume that after excitation there is no coherence between the spins. In other words,  the time scale at which the spins dephase is the shortest time scale within the system, such that the density matrix  $\hat{\rho}^S$ is diagonal. The diagonal elements of $\hat{\rho}^S$ are given by the occupation numbers $\rho^S_{m_s m_s}=f_{m_s}$ for each energy level $m_s$ of a single spin, where $m_s$ corresponds to the $z$ component of the spin. In this Boltzmann approach, the ensemble average of the spin in the $z$ direction is given by $\langle\hat{S}^{d,z}\rangle = \sum_{m_s=-S}^S m_s f_{m_s} $. 

In order to find the magnetization dynamics, we calculate the time derivative of all occupation numbers $f_{m_s}$. The mathematical description follows from the Liouville-von Neumann equation, and a coarse-grained description of the time evolution of the density operator \cite{Iotti2005}. The coarse-graining step size, interval $\delta t$, determines the time resolution of the model and should be sufficiently small compared to the observed demagnetization time $\tau_M$. Moreover, we assume that the time interval $\delta t$ satisfies the conditions for the Markov approximation, i.e., $\delta t$ should be much larger than the correlation time of the electrons and the density matrix changes relatively slowly \cite{Cywinski2007,Nieves2014}. Secondly, it is assumed that the time interval is much larger than the time scale associated with the energy transfer, in this case that yields $\delta t \gg \hbar/\Delta $ \cite{Nieves2014}.  This is the standard limit underlying Fermi's golden rule, i.e., the condition leads to the transitions having a well-defined energy conservation represented by the Dirac delta function. Hence, we should have that $\hbar/\Delta  \ll \delta t \ll \tau_M$. Since $\hbar/\Delta \sim 10\mbox{ fs}$ (having $\Delta\sim k_B T_C$ and Curie temperature $T_C\sim 1000\mbox{ K}$) and $\tau_M$ is of the order of $\sim 100\mbox{ fs}$, the  validity of this limit is not trivial.  However, it is expected that the  role of all the approximations is relatively weak and only affects the results quantitatively.  

Finally, using the diagonality of the density matrix $\hat{\rho}^S$ and the explicit form of the interaction Hamiltonian $\hat{H}_{sd}$, the rate equation can be written as \cite{Cywinski2007}

\begin{eqnarray}
\label{eq:eq4}
\dfrac{df_{m_s}}{dt} &=& -(W_{m_s-1,m_s} +W_{m_s+1 ,m_s} ) f_{m_s} 
\\
\nonumber 
&& +W_{m_s, m_s-1} f_{m_s-1} + W_{m_s,m_s+1} f_{m_s+1},
\end{eqnarray}

\noindent where $W_{m_s\pm 1,m_s}$ are the transition rates from level $m_s$ to $m_s\pm 1$. The transition rates are calculated using Fermi's golden rule, analogous to the derivation of the M3TM \cite{Koopmans2010}. We assume that the $s$ electrons thermalize rapidly due to Coulomb scattering and can be described by Fermi-Dirac statistics. Here, the distributions for the spin up and spin down $s$ electrons have a common temperature $T_e$, but are allowed to have a distinct chemical potential for which the difference is defined as the spin accumulation $\mu_s = \mu_\uparrow-\mu_\downarrow$ \cite{Tveten2015,Cywinski2007}. In the limit that the Fermi energy is much larger than all other energy scales, the transition rate is given by \cite{Cywinski2007,Gridnev2016}

\begin{eqnarray}
\label{eq:eq5}
W_{m_s\pm 1,m_s} &=& \dfrac{\pi}{2 \hbar} J_{sd}^2 S^{\pm}_{m_s} D_{\uparrow}D_{\downarrow} (\Delta-\mu_s)
\dfrac{ \exp \bigg( \mp \dfrac{\Delta-\mu_s}{2 k_B T_e}  \bigg) }{2 \sinh{
\bigg(\dfrac{\Delta-\mu_s}{2 k_B T_e}\bigg)}}.
\end{eqnarray}

\noindent Here, $S^{\pm}_{m_s} = S(S+1) - m_s(m_s\pm 1)$ and $D_{\uparrow,\downarrow} $ (in units $\mbox{eV}^{-1}\mbox{atom}^{-1} $) is the density of states at the Fermi level for the spin up and spin down $s$ electrons respectively. Equation (\ref{eq:eq5}) mathematically quantifies the amount of available phase space for transitions induced by the $s$-$d$ interaction. Figures \ref{fig:fig1}(b)-\ref{fig:fig1}(d) schematically show the changes to the occupation of the $d$ and $s$ electron states as a result of laser heating the system. Figure \ref{fig:fig1}(c) shows that the thermal broadening of the Fermi-Dirac functions allows for transitions between the two spin directions of the $s$ electrons, which is accompanied by a flip of a localised $d$ electron spin. The $s$ electrons thermalize rapidly and the new distributions have a shifted chemical potential, i.e., a non-zero spin accumulation is created, as is depicted in Fig. \ref{fig:fig1}(d). 

The dynamics of the spin accumulation $\mu_s$ can be derived analogously and directly follows from spin angular momentum conservation. Now we define the normalized magnetization $m_d =-\langle \hat{S}^{d,z} \rangle/S $ of the localised magnetic moments. In equilibrium,  the experimentally detectable magnetization is dominated by $m_d$. This is not straightforward after excitation because of the induced exchange of angular momentum between the $s$ and $d$ electrons. In general, the magneto-optical signal in typical pump-probe experiments is a linear superposition of the contribution of the $s$ and $d$ electrons. For a $S=1/2$ system we have $\Delta=2k_B T_C m_d$,  and the dynamics is described by the two equations

\begin{eqnarray}
\label{eq:eq6}
\dfrac{d \mu_s}{ d t} &=& \rho_{sd}   \dfrac{dm_d}{dt} -\dfrac{\mu_s}{\tau_{s}} ,
\\
\label{eq:eq7}
\dfrac{dm_d}{dt} &=& \dfrac{1}{\tau_{sd} } \Big(m_d-\dfrac{\mu_s}{2 k_B T_C} \Big)
\bigg[ 1-m_d \coth{ \Big(\dfrac{2m_d k_B T_C - \mu_s}{2 k_B T_e} \Big) }
\bigg] ,
\end{eqnarray}

\noindent where we used the definition $\rho_{sd}= \bm\bar{D}^{-1} - J_{sd}/2$, with $\bm\bar{D}=2 D_{\uparrow}D_{\downarrow}/(D_{\uparrow}+D_{\downarrow})$ from Ref.\ \cite{Tveten2015}.  Note that the term proportional to $J_{sd}$ results from the energy gap between the spin up and spin down $s$ electrons arising from the $s$-$d$ interaction (as was introduced in the transition from Eq. (\ref{eq:eq2}) to Eq.  (\ref{eq:eq3})), which can be both positive and negative depending on the sign of $J_{sd}$. Moreover, we defined $\tau_{sd}^{-1}= (\pi/\hbar) J_{sd}^2 D_\uparrow D_\downarrow k_B T_C$, which is closely related to the demagnetization rate. We introduced the phenomenological term proportional to $\tau_\mathrm{s}^{-1}$, which describes all spin-flip scattering processes that dissipate angular momentum out of the combined electronic system \cite{Tveten2015}, e.g., this term includes Elliott-Yafet electron-phonon scattering.

Equation (\ref{eq:eq7}) clearly shows the similarities with the standard form of the equation for the longitudinal magnetization relaxation of a spin $S=1/2$ system within a mean-field approach. For instance, in the limit $\tau_s\rightarrow 0$ the spin accumulation directly vanishes and the equilibrium condition is given by $m_d=\tanh( m_d T_C/T_e) $. In this limit there is no net spin polarization, i.e., the $s$ electrons can be considered as spinless, which is exactly the assumption underlying the M3TM \cite{Koopmans2010}. We note that although this expression closely resembles the expression presented in Ref.\ \cite{Koopmans2010}, the prefactor corresponds to a completely different physical mechanism. More details about the relation with the M3TM will be discussed below. 

Although we have a simple definition of the parameters $\rho_{sd}$ and $\tau_{sd}$, the estimation of these parameters is far from straightforward. We approximated the $d$ and $s$ electrons as two distinct systems, localised and itinerant electrons. In the real system there is no such clear separation because of $s$-$d$ hybridization. Effectively, we separated the `band-like' and `local magnetic' properties of the combined electronic system ($d$ and $s$), which makes it complex to estimate the relevant value of $\bm\bar{D}$. Hence, it is convenient to treat both $\tau_{sd}$ and $\rho_{sd}$ as effective parameters. In the upcoming sections, we interpret $\tau_{sd}$ as the experimentally retrieved demagnetization time and we choose the constant $\rho_{\mathrm{sd}}=1\mbox{ eV}$.  The exact values should be retrieved from carefully fitting the model to the experiments, which is beyond the scope of this theoretical paper.

Finally, $\bar{D}^{-1}$ ($D_{\uparrow,\downarrow}^{-1}$) scales with the width of the conduction band and is typically much larger than $J_{sd}/2$, i.e., we have $\rho_{sd}=\bar{D}^{-1}-J_{sd}/2 \approx \bar{D}^{-1}$ . Then, we can define the  magnetization of the total spin system ($s$ and $d$ electrons) as $m_{\mathrm{tot}} = m_d - \rho_{sd}^{-1} \mu_s $, which is  conserved by the $s$-$d$ interaction and will be used in the following analyses.

In the next section we discuss the important role of the spin accumulation by describing the laser-induced demagnetization experiments using the numerical solutions of Eqs. (\ref{eq:eq6}) and (\ref{eq:eq7}). 

\begin{figure}[ht!]
\includegraphics[scale=0.95]{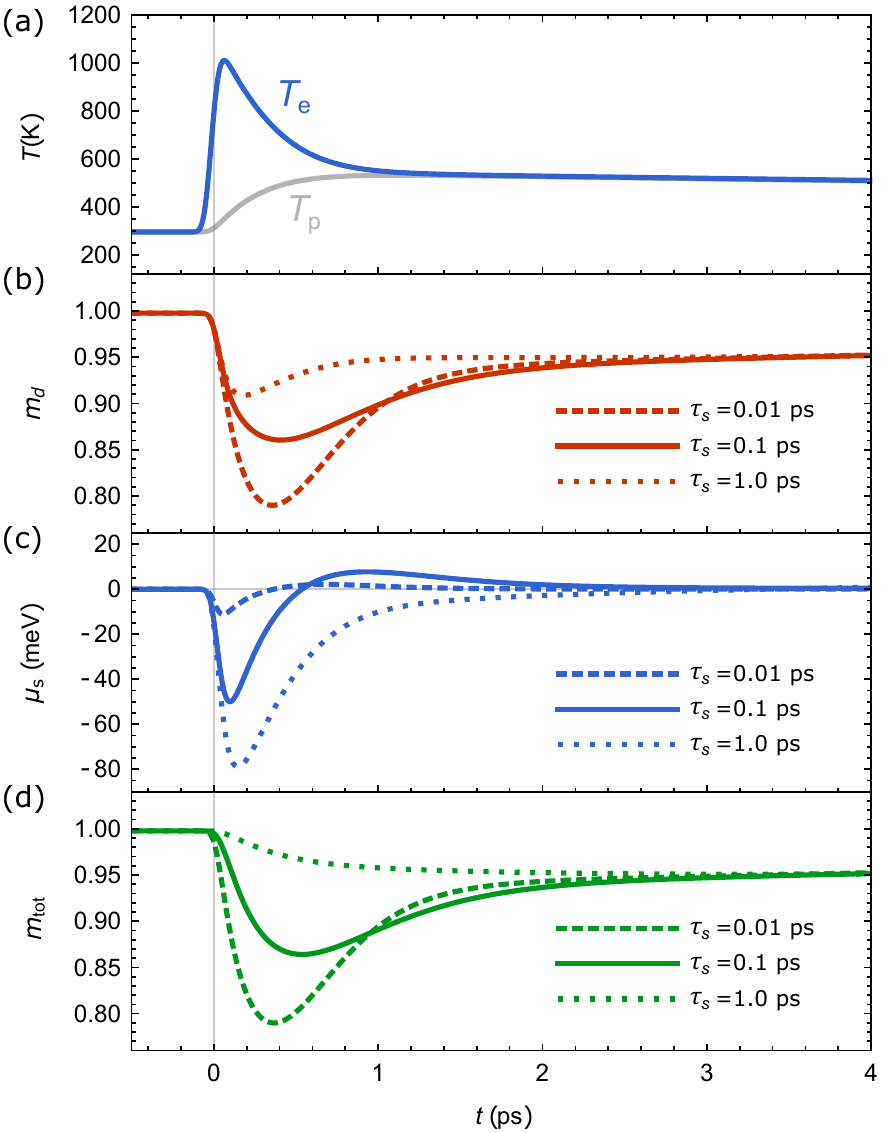}
\caption{\label{fig:fig2} Ultrafast demagnetization described by the  $s$-$d$ model. Figure (a) shows the temporal profile of the $s$ electron temperature $T_e$ and phonon temperature $T_p$, after laser-pulse excitation at $t=0$ with $P_0 = 12\cdot 10^8\mbox{ Jm}^{-3}$. (b)-(d) Present the laser-induced dynamics of the spin systems, using $T_C=1000\mbox{ K}$ and $\tau_{sd} = 0.2 \mbox{ ps}$. Here, the line types indicate the calculations for different values of $\tau_s$, which are given in the figure. (b) Shows the resulting magnetization dynamics in the $d$ electron system. (c) Shows the temporal profile of the spin accumulation $\mu_s$ and (d) shows the dynamics of the total magnetization $m_{\mathrm{tot}}$. 
} 
\end{figure}

\section{III.$\qquad$ Ultrafast demagnetization} 

In order to investigate the typical laser-induced dynamics of the local magnetization and spin accumulation specifically, we consider a system with magnetic parameters $\tau_{sd}=0.2\mbox{ ps}$ and $T_C=1000\mbox{ K}$. To model the laser heating we define the temporal profile of the laser pulse as $P(t)=(P_0/(\sigma \sqrt{\pi}))\exp [-(t-t_0)^2/\sigma^2]$, where $P_0$ is the absorbed laser pulse energy density and $\sigma$ determines the pulse duration, which is set to  $50\mbox{ fs}$. We use the standard two-temperature model to find the dynamics of the $s$ electron temperature $T_e$ and phonon temperature $T_p$ \cite{Anisimov1974}. The two-temperature model describes the equilibration of $T_e$ and $T_p$ by electron-phonon scattering. We include a heat dissipation term that transfers heat out of the phonon system on a time scale $\tau_D=20\mbox{ ps}$. For the heat capacities and the electron-phonon coupling constant we use the values for cobalt given in Ref.\ \cite{Koopmans2010}. We calculate the dynamics of the magnetization and spin accumulation by solving Eqs. (\ref{eq:eq6}) and (\ref{eq:eq7}) numerically. We do this for multiple values of $\tau_\mathrm{s}$. The results are presented in Figs. \ref{fig:fig2}(a)-(d).

Figure \ref{fig:fig2}(a) shows the laser heating of the $s$ electrons and the equilibration of the electron temperature with the phonon temperature. Figures \ref{fig:fig2}(b)-(d) display the laser-induced  dynamics of the spin systems for different values of the spin-flip scattering time $\tau_s$, as indicated by the different line types. Figure \ref{fig:fig2}(b) shows the magnetization of the $d$ electrons $m_d$ as a function of time. The temporal profile of the spin accumulation $\mu_s$ is presented in Fig. \ref{fig:fig2}(c). Finally, Fig. \ref{fig:fig2}(d) displays the total magnetization $m_{\mathrm{tot}}$ as a function of time. Figures \ref{fig:fig2}(b) and \ref{fig:fig2}(d) clearly show that the demagnetization of $m_d$ and $m_{\mathrm{tot}}$ is maximized for the smallest $\tau_s$.

\begin{figure}[ht!]
\includegraphics[scale=0.95]{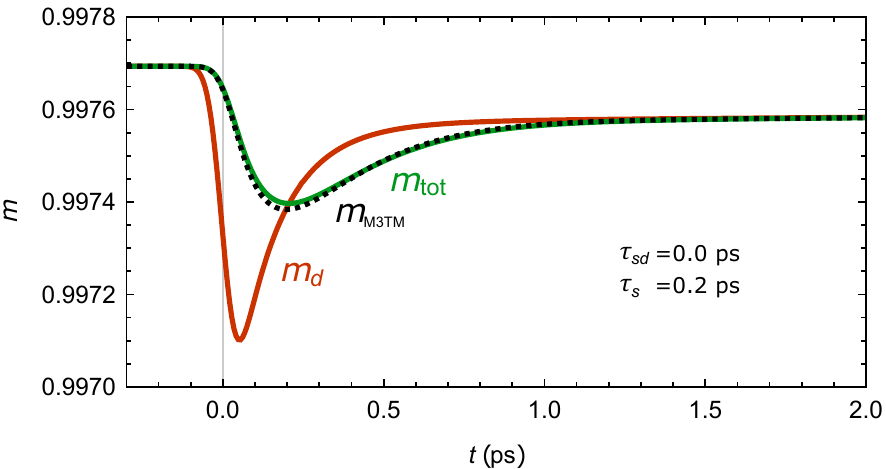}
\caption{\label{fig:fig3} The $s$-$d$ model in the limit of a strong $s$-$d$ coupling ($\tau_{sd}\rightarrow0$), compared to the microscopic three-temperature model (M3TM) \cite{Koopmans2010}. The plot shows the magnetizations $m_d$ (red) and $m_\mathrm{tot}$ (green) as a function of time after laser-pulse excitation at $t=0$ with  $P_0=0.1\cdot 10^8\mbox{ Jm}^{-3}$ and $\sigma=50\mbox{ fs}$. The remaining magnetic parameters are given by $T_C=1000\mbox{ K}$ and $\tau_s=0.2 \mbox{ ps}$. The dotted black line indicates the magnetization calculated with the basic M3TM (using demagnetization time scale $\tau_M=\tau_s=0.2\mbox{ ps}$ \cite{Footnote}). 
} 
\end{figure}

The calculations show that the creation of a spin accumulation has a negative feedback effect on the demagnetization (of both $m_d$ and $m_{\mathrm{tot}}$), i.e., the short-lived spin accumulation acts as a bottleneck \cite{Cywinski2007,Tveten2015}. The bottleneck can be removed by the additional spin-flip relaxation processes in the $s$ electron system, which happen at a rate given by $\tau_s^{-1}$. This means that in the limit $\tau_{sd}\ll \tau_{s}$ the demagnetization rate  strongly depends on $\tau_{s}$. In the extreme case $\tau_{sd}\rightarrow 0$, which corresponds to an infinitely strong $s$-$d$ interaction, $m_d$ and $\mu_s$ are equilibrated instantaneously and their relation can be found by setting Eq. (\ref{eq:eq7}) equal to zero. Now the $d$ and $s$ electrons can be treated as a single spin system with magnetization $m_\mathrm{tot}$ of which the subsequent dynamics is governed by $T_e$ and the additional spin-flip scattering processes of the $s$ electrons. These additional scatterings include Elliott-Yafet electron-phonon scattering. Hence, in analogy with the M3TM \cite{Koopmans2010}, the system behaves as a single spin system with a characteristic demagnetization rate that is associated with Elliott-Yafet electron-phonon scattering. More specifically, in the low-fluence limit and having temperatures well below the Curie temperature, $m_\mathrm{tot}(t)$ converges to the magnetization dynamics from the M3TM, which is visualized in Fig. \ref{fig:fig3}. Here,  $m_d(t)$ and $m_\mathrm{tot}(t)$ follow from the $s$-$d$ model using $P_0=0.1\cdot 10^8\mbox{ Jm}^{-3}$ and $\tau_s=0.2 \mbox{ ps}$ in the limit of a strong $s$-$d$ coupling ($\tau_{sd}\rightarrow 0$). All other system parameters are kept equal to the calculations of Fig. \ref{fig:fig2}. The dotted black line is the magnetization described by the M3TM for the same system, using the demagnetization time scale $\tau_M=\tau_s=0.2\mbox{ ps}$ \cite{Footnote}, which shows a clear overlap with the total magnetization $m_\mathrm{tot}$.  

On the other hand, in the limit $\tau_{sd}\gg\tau_{s}$ the spin accumulation relaxes efficiently and the bottleneck effect is negligible. In this limit, the magnetizations $m_d$ and $m_{\mathrm{tot}}$ converge and their dynamics can be well described by Eq. (\ref{eq:eq7}) without the terms involving $\mu_s$ (similar to the limit $\tau_{s}\rightarrow 0$). Up to a prefactor, the magnetizations $m_d$ and $m_{\mathrm{tot}}$ are now described by the same mathematical expression as in the M3TM. However, the physical origin of the ultrafast demagnetization is different. 

In conclusion, in both regimes there is a clear relation with the M3TM. Nevertheless, in a real system it is expected that $\tau_{sd}$ and $\tau_{s}$ can be of the same order and a short-lived spin accumulation influences the magnetization dynamics. Finally, Fig. \ref{fig:fig2}(c) shows that for a decreasing $\tau_{s}$ the spin accumulation becomes directly proportional to the temporal derivative of the magnetization $m_d$, as can be mathematically derived from Eq. (\ref{eq:eq6}) in the limit $\tau_{s}\rightarrow 0$. These typical curves for $\mu_s$ resemble the measurements in the experimental investigations of the optically generated spin currents at the interface of a ferromagnetic and non-magnetic metal \cite{Choi2014}.

In the following sections we investigate the role of spin transport on the demagnetization process.

\section{IV.$\qquad$ F/N/F structures: parallel versus antiparallel}

\begin{figure}[t!]
\includegraphics[scale=0.95]{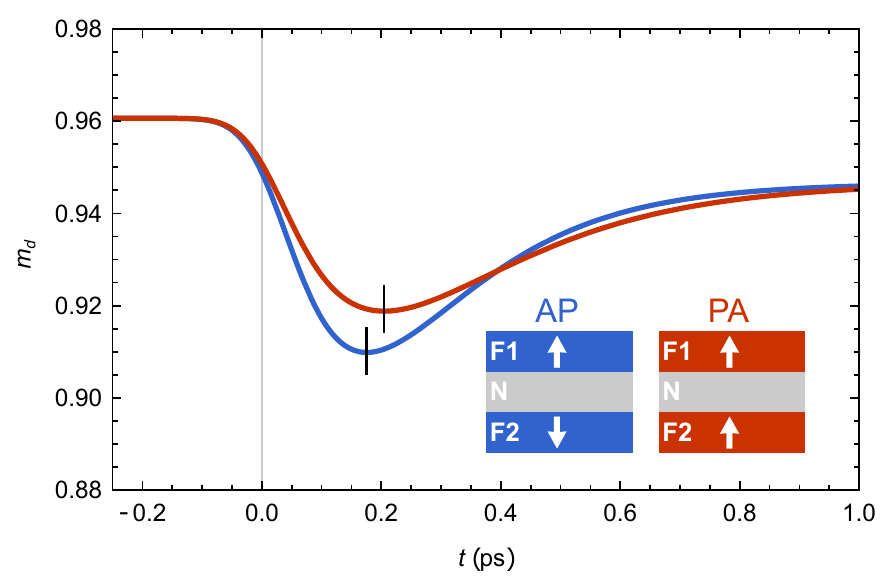}
\caption{\label{fig:fig4} The laser-induced magnetization dynamics of the F/N/F structure in the antiparallel and parallel configurations, described by the $s$-$d$ model. We used a low-energetic laser pulse with $P_0=1\cdot 10^8\mbox{ Jm}^{-3}$ and $\sigma = 70\mbox{ fs}$. The diagram shows the magnetization $m_d$ of F layer 1 as a function of time. The color scheme indicates the specific configuration, as is indicated in the inset. The F layers have magnetic parameters $T_C=600\mbox{ K}$, $\tau_{sd}=0.1\mbox{ ps}$ and $\tau_{s}=0.02\mbox{ ps}$. 
} 
\end{figure}

In the previous section we showed that during laser-pulse excitation  a spin accumulation is generated that counteracts the demagnetization process. In this section, we show that spin transport can act as an additional mechanism for removing this bottleneck effect. We model the experiments with collinear magnetic heterostructures \cite{malinowski2008control,Rudolf2012}. More specifically, we address the results presented in Ref.\ \cite{malinowski2008control}, in which a magnetic heterostructure is investigated that consists of two identical Co/Pt multilayers separated by a Ru spacer layer. The authors present a comparison of the demagnetization of the parallel and antiparallel aligned states of the heterostructure. The measurements showed that the antiparallel configuration has a larger demagnetization rate and amplitude, which can be explained by the generation of a spin current that enhances the demagnetization process. In the following, we will show that these results can be understood and reproduced by the presented $s$-$d$ model. 

Hence, we consider a system containing two identical ferromagnetic (F) layers with a non-magnetic (N) layer in between.  We further refer to this system as the F/N/F structure. We investigate the different laser-induced demagnetization rates for the parallel and antiparallel configuration of the F/N/F structure. The systems are schematically depicted in the inset of Fig. \ref{fig:fig4}. By definition, F layer 1 is pointing up in both configurations, whereas F layer 2 is pointing in the up and down directions for the parallel and antiparallel configurations respectively. 

We assume all the layers to be very thin, such that we can take the temperature, magnetization and spin accumulation homogeneous within each layer. We define a magnetization $m_{d,i}$ and $\mu_{s,i}$ for each F layer $i$. Because of the very small thickness of the N layer we assume that the electron transport is in the ballistic regime.  In that case, we can approximate that the spin transport in the non-magnetic layer is purely driven by the difference in the spin accumulation of both F layers. Within these limits the spin accumulations satisfy

\begin{eqnarray}
\label{eq:eq8}
\dfrac{d \mu_{s,i} }{ d t} &=& \rho_{sd}  \dfrac{dm_{d,i}}{dt} -\dfrac{\mu_{s,i}}{\tau_{\mathrm{s},i}} -\dfrac{\mu_{s,i}-\mu_{s,j}}{\tau_{\mathrm{B}}} ,
\end{eqnarray}

\noindent where $i\neq j$ and $i,j\in \{1,2\} $. The last term, which is introduced phenomenologically,  represents the spin transfer between the F layers driven by ballistic electron transport and enforces the spin accumulations to equilibrate. The prefactor is defined in terms of the time scale $\tau_B$. We use that $\tau_B\sim 1\mbox{ fs}$ based on the assumptions that the Fermi velocity is $v_F\sim 10^6\mbox{ ms}^{-1}$ and the thickness of the N layer is $d_N \sim 1\mbox{ nm}$. Note that the transport term depends on the spin accumulation at the same time coordinate, i.e., the distinct F layers feel changes in the opposing layer instantaneously. In the real experiment there might be a small delay. However, we expect that this effect can be neglected in our calculations. Finally, we stress that this particular form of the transport term can only be used for two strictly identical F layers, as was the case in Ref.\ \cite{malinowski2008control}.

For the F layers we use the magnetic parameters $\tau_{\mathrm{sd}}=0.1\mbox{ ps}$, $\tau_{\mathrm{s}}=0.02\mbox{ ps}$ and $T_C=600\mbox{ K}$, which are approximated values corresponding to the Co/Pt multilayers used in the experiments \cite{malinowski2008control}. Furthermore, we apply a low-energetic laser pulse with $P_0=1 \cdot 10^8 \mbox{ Jm}^{-3}$ and assume that the system is heated homogeneously. In this specific case, we set the pulse duration to $\sigma=70\mbox{ fs}$ \cite{malinowski2008control}. For convenience, we still use the heat capacities and electron-phonon coupling constant of pure cobalt \cite{Koopmans2010}. 

The results are displayed in Fig. \ref{fig:fig4}. The red and blue curves show the magnetization $m_d$ of F layer 1 for the parallel and antiparallel configuration respectively. It is verified that $m_{\mathrm{tot}}$ (not shown) behaves very similar. In agreement with the experiments, we observe a larger demagnetization rate and amplitude for the antiparallel configuration. This can be easily understood from the transport term in Eq. (\ref{eq:eq8}). In the parallel configuration we have $\mu_{s,1}=\mu_{s,2}$ at any time  and the transport term vanishes. In contrast, for the antiparallel configuration we have $\mu_{s,1}=-\mu_{s,2}$, the transport does not vanish and behaves as an extra channel for angular momentum transfer. This extra channel assists the reduction of the spin accumulation, thereby leading to a larger demagnetization. Equivalently, in the antiparallel configuration the spin current in the non-magnetic layer is non-zero and has exactly the correct polarization to enhance the demagnetization rates in both F layers. Finally, Fig. \ref{fig:fig4}  also shows that the demagnetization curves of the two configurations converge at $t\sim 400\mbox{ fs}$, which is in agreement with the experiments \cite{malinowski2008control}.

In the next section, we analyze the temporal profile of the spin current generated in an F/N structure in the diffusive regime. Furthermore, we investigate the role of the thickness of the layers. 

\begin{figure}[t!]
\includegraphics[scale=0.95]{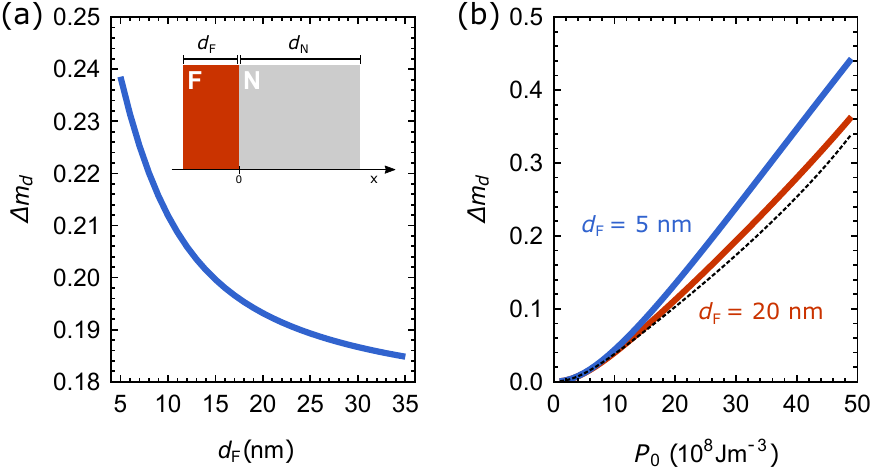}
\caption{\label{fig:fig5} The laser-induced magnetization dynamics in an F/N structure. The magnetic parameters of the ferromagnetic layer are given by $\tau_{sd}=0.3\mbox{ ps}$, $\tau_{s}=0.2\mbox{ ps}$ and $T_C=1388\mbox{ K}$. (a) The maximum demagnetization $\Delta m_d$ (averaged over the F layer) as a function of the ferromagnetic layer thickness $d_F$. The non-magnetic layer thickness is set to $d_N=200\mbox{ nm}$ and we used $P_0 = 30\cdot 10^8\mbox{ Jm}^{-3}$ . The inset shows the system schematically. (b) The maximum demagnetization $\Delta m_d$ as a function of $P_0$ for $d_F=5\mbox{ nm}$ (blue) and $d_F=20 \mbox{ nm}$ (red). For both systems we have $d_N=200\mbox{ nm}$. The black dashed line indicates the demagnetization of the bulk ferromagnet in the absence of an F/N interface ($d_F=\infty$ and $d_N=0$).  
} 
\end{figure}

\section{V.$\qquad$ F/N structures: diffusive spin transport}

Finally, we introduce diffusive spin transport in the simplified $s$-$d$ model and we show that in magnetic heterostructures spin diffusion within the $s$ electron system can significantly enhance the demagnetization rate. Here, we model a system consisting of a ferromagnetic (F) layer and a non-magnetic (N) layer. In contrast to the similar approach reported in Refs.\ \cite{Choi2014,Shin2018,Kimling2017}, we  calculate the local magnetization dynamics $dm_d/dt$ directly from the $s$-$d$ model using Eq. (\ref{eq:eq7}), that serves as a source for spin-polarized $s$ electrons via Eq. \ref{eq:eq6}. Thereby, we can specifically address the mutual influence of the dynamics of the local magnetization $m_d$ and spin accumulation $\mu_s$ both as a function of position and time.

As indicated in the inset of Fig. \ref{fig:fig5}(a), we define the thickness of the F layer and N layer as $d_F$ and $d_N$ respectively. Spin transport is described in the diffusive regime, where both layers are treated on an equal footing \cite{Slachter2010}. For convenience, we assume that the system is heated homogeneously, i.e., there are no thermal gradients present. Hence, the demagnetization of the F layer is the only source of the spin current and there is no spin-dependent Seebeck effect included in this calculation. 

It is assumed that that the interface is transparent for spins, such  that the spin accumulation  is continuous at the interface. Imposing that there is no charge transport, the spin current density can be expressed as $j_s=-(\bar{\sigma}/e^2)\partial \mu_s/\partial x $ \cite{Slachter2010,Kimling2017}, with $\bar{\sigma}=2\sigma_\uparrow\sigma_\downarrow/(\sigma_\uparrow+\sigma_\downarrow)$ and $\sigma_{\uparrow,\downarrow}$ the spin-dependent electrical conductivity. Combining this with the continuity equations for spin-up and -down electrons  \cite{Beens2018,Kimling2017}, gives that introducing diffusive spin transport in Eq. (\ref{eq:eq6}) leads to  \cite{Kimling2017}

\begin{eqnarray}
\label{eq:eq9}
\dfrac{\partial \mu_s}{\partial t}
-\dfrac{1}{\bar{\nu}} \dfrac{\partial }{\partial x}
\bigg[\dfrac{\bar{\sigma}}{e^2} 
\dfrac{\partial \mu_s}{\partial x} 
\bigg]
&=&
\rho_{sd} \dfrac{\partial m_d}{\partial t} -\dfrac{\mu_s}{\tau_s} ,
\end{eqnarray}

\noindent where  all variables explicitly depend on the spatial coordinate $x$ and we assumed that the system is homogeneous in the lateral directions. The interface is at $x=0$.  Here, $\bar{\nu}=2\nu_\uparrow\nu_\downarrow/(\nu_\uparrow+\nu_\downarrow)$ with $\nu_{\uparrow,\downarrow}$ the spin-dependent density of states in units per volume per energy ($\bar{\nu}=\bar{D}/ V_\mathrm{at}$). The N layer ($x>0$) is characterised by $\sigma_\uparrow=\sigma_\downarrow$ and $\nu_\uparrow=\nu_\downarrow$, and the absence of the ($s$-$d$ interaction) source term. Equation (\ref{eq:eq9}) is further simplified by imposing that $\bar{\nu}$ is independent of $x$, where we have assumed that in the F layer $\nu_\uparrow\sim\nu_\downarrow$ and $\sigma_\uparrow\sim\sigma_\downarrow$. These choices are made for convenience and are consistent with the underlying assumptions of the $s$-$d$ model, that the $d$ electrons in the F layer do not contribute to the conductive properties of the material. In that case,  Eq. (\ref{eq:eq9}) reduces to

\begin{eqnarray}
\label{eq:eq10}
\dfrac{\partial \mu_{s}}{\partial t} &=&
\rho_{sd} \dfrac{\partial m_d}{\partial t} 
  -   \dfrac{\mu_{s}}{\tau_{s}}+\dfrac{\partial}{\partial x}\bigg[ D_\mathrm{diff} \dfrac{\partial \mu_{s}}{\partial x} \bigg] .
\end{eqnarray}

\noindent  with the diffusion coefficient  $D_{\mathrm{diff}} =\bar{\sigma}/(\bar{\nu} e^2)$ \cite{Shin2018,Kimling2017}. Finally, we set the spin currents at the edges equal to zero $j_s(-d_F)=j_s(d_N)= 0$. 

Equation (\ref{eq:eq10}) is solved numerically, where we discretized the system using a finite difference method. Note that the spatial derivative of the diffusion coefficient $D_\mathrm{diff}$ is only non-zero at the interface. In these calculations, the F layer corresponds to pure cobalt  for which we use the diffusion coefficient $D_\mathrm{diff}=250\mbox{ nm}^2\mbox{ps}^{-1}$\cite{Shin2018}. Furthermore, we use  $\tau_{sd}=0.3 \mbox{ ps}$ and $\tau_{s}=0.2\mbox{ ps}$.  For the N layer we take $D_\mathrm{diff}=9500\mbox{ nm}^2\mbox{ps}^{-1}$ and $\tau_s=25\mbox{ ps}$, which correspond to the diffusion coefficient and spin-flip relaxation time for Copper \cite{Shin2018}. 

Figure \ref{fig:fig5}(a) shows a calculation of the F/N structure excited with a laser pulse with energy density $P_0 =30\cdot 10^8\mbox{ Jm}^{-3}$ and pulse duration $\sigma=50\mbox{ fs}$, where we used the heat capacities and electon-phonon coupling constant of cobalt \cite{Koopmans2010}. The diagram shows the maximum demagnetization $\Delta m_d=m_{d,0}-m_{d,\mathrm{min}}$ (averaged over the F layer) as a function of the F layer thickness $d_F$, where $m_{d,0}$ is the initial (equilibrium) value of $m_d$, and $m_{d,\mathrm{min}}$ is the minimum of $m_d$ after excitation. The thickness of the N layer is kept constant and set to $d_N=200\mbox{ nm}$. It clearly shows that the demagnetization becomes larger when the F layer thickness decreases. Intuitively, the injection of spins into the non-magnetic layers can enhance the demagnetization significantly as long as the F layer is relatively thin. This conclusion is corroborated by the results presented in Fig. \ref{fig:fig5}(b), which shows the demagnetization $\Delta m_d $ as a function of $P_0$. The results are plotted for $d_F=5\mbox{ nm}$ and $d_F=20\mbox{ nm}$. The dashed line indicates the demagnetization of a bulk ferromagnet in the absence of an N layer ($d_F=\infty$ and $d_N=0$). The calculations show that for a relatively thin F layer spin injection into the N layer can lead up to $\sim 30\%$ more demagnetization.

\begin{figure}[t!]
\includegraphics[scale=0.95]{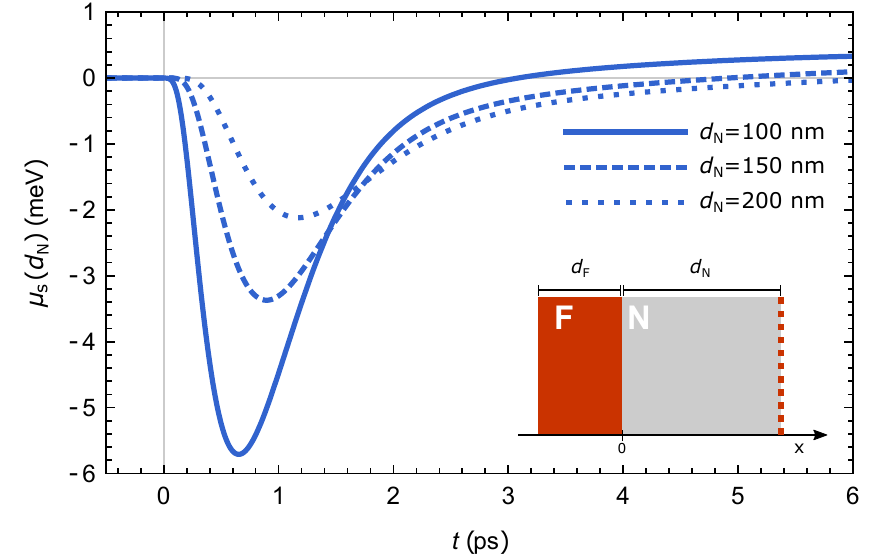}
\caption{\label{fig:fig6} The diffusive spin current injected in the non-magnetic (N) layer with thickness $d_N$, coupled to a ferromagnetic (F) layer with thickness $d_F$. The diagram shows the spin accumulation at the outer edge of the N layer (position $x=d_N$, indicated by the red dotted line in the inset) as a function of time. Furthermore, we set $P_0=20\cdot 10^8\mbox{ Jm}^{-3} $ and $\sigma=50\mbox{ fs}$. The magnetic parameters are identical to the values used in Fig. \ref{fig:fig5}. The line types indicate the results for three different values of $d_N$. The thickness of the F layer is kept constant and set to  $d_F=10\mbox{ nm}$.  
} 
\end{figure}

Now we discuss the dynamics of the injected spin current itself. We do this by calculating the spin accumulation at the outer edge of the N layer $\mu_s(d_N)$. The results are shown in Fig. \ref{fig:fig6}, which displays the spin accumulation as a function of time for three different values of $d_N$ that are given in the figure. The F layer thickness is kept constant at $d_F=10\mbox{ nm}$. In agreement with the  experimental investigations \cite{Choi2014,Shin2018}, the diagram clearly shows that for an increasing $d_N$ the minimum of $\mu_s(d_N)$ shifts in time and is reduced. This behavior can be understood from the diffusive character of the spin transport. Here, the temporal profile of $\mu_s(d_N)$ is highly sensitive to the specific material that composes the F layer and the corresponding effective parameters, as is expected from the experimental and numerical investigation using various materials for the F layer \cite{Shin2018}.   

A more quantitative comparison with the experiments would require addressing spin transport beyond the diffusive regime and implementing a finite penetration depth of the laser pulse in the modeling. However, we focused our discussion on the dynamics that stems from the $s$-$d$ interaction and we specifically investigate the role of $\mu_s$ independent of the thermal properties of the system. In that case, the model shows that in the presence of only the $s$-$d$ interaction, the typical experimental observations can be explained and show a qualitative agreement \cite{Choi2014,Shin2018}.

\section{VI.$\qquad$ Conclusion and discussion}

In conclusion, we discussed a simplified $s$-$d$ model that we used to describe laser-induced magnetization dynamics in magnetic heterostructures and to study the interplay between local and nonlocal spin dynamics. The presented numerical calculations emphasize the critical role of the spin accumulation. During demagnetization a spin accumulation is created, which counteracts the demagnetization process. Both local spin-flip scatterings and spin transfer to a non-magnetic layer can reduce this spin accumulation effectively and, depending on the system, can both play a dominant role in the characterization of the demagnetization rate. Importantly, the modeling shows that even in the absence of any other interaction, the $s$-$d$ interaction could account for the typically observed ultrafast phenomena, such as being a driving force for laser-induced spin transport in magnetic heterostructures.

The presented analyses show that the simplified $s$-$d$ model provides a versatile description of ultrafast magnetization dynamics, which converges to the M3TM for a strong $s$-$d$ coupling and possesses the additional feature that spin transport can be included straightforwardly. However, one needs to keep in mind that the model is a simplified description of the underlying physics. As was earlier discussed, the $d$ electrons are not perfectly localised.  Moreover, the $d$ electron spins are described using a Weiss model, i.e., spin wave excitations are  neglected. In a more complete description, the $d$ electrons are described as a magnonic system and the $s$-$d$ interaction corresponds to electron-magnon scattering \cite{Tveten2015}. That has the advantage that spin transport driven by magnon transport can be included, which is expected to give a non-negligible contribution to the spin transport at the interface between a ferromagnetic and non-magnetic metal \cite{Beens2018}. In that case, the electronic and magnonic contribution to the spin transport can be treated on an equal footing by introducing a magnon chemical potential \cite{cornelissen2016magnon,Beens2018}. This description should allow for both chemical potential gradients and thermal gradients. For instance, thermal gradients can be induced by a finite penetration depth of the laser pulse and can drive a spin current via the electronic spin-dependent Seebeck effect \cite{Choi2015,Alekhin2017} and the magnonic spin Seebeck effect \cite{Uchida2010,Xiao2010}. Nevertheless, we expect that the dominant contributions to the dynamics can be well described by the simplified  $s$-$d$ model including spin transport and it provides a useful pathway to investigate the underlying physics.

This work is part of the research programme of the Foundation for Fundamental Research on Matter (FOM), which is part of the Netherlands Organisation for Scientific Research (NWO). R.D. is member of the D-ITP consortium, a program of the NWO that is funded by the Dutch Ministry of Education, Culture and Science (OCW). This work is funded by the European Research Council (ERC).


\begin{thebibliography}{37}%
\makeatletter
\providecommand \@ifxundefined [1]{%
 \@ifx{#1\undefined}
}%
\providecommand \@ifnum [1]{%
 \ifnum #1\expandafter \@firstoftwo
 \else \expandafter \@secondoftwo
 \fi
}%
\providecommand \@ifx [1]{%
 \ifx #1\expandafter \@firstoftwo
 \else \expandafter \@secondoftwo
 \fi
}%
\providecommand \natexlab [1]{#1}%
\providecommand \enquote  [1]{``#1''}%
\providecommand \bibnamefont  [1]{#1}%
\providecommand \bibfnamefont [1]{#1}%
\providecommand \citenamefont [1]{#1}%
\providecommand \href@noop [0]{\@secondoftwo}%
\providecommand \href [0]{\begingroup \@sanitize@url \@href}%
\providecommand \@href[1]{\@@startlink{#1}\@@href}%
\providecommand \@@href[1]{\endgroup#1\@@endlink}%
\providecommand \@sanitize@url [0]{\catcode `\\12\catcode `\$12\catcode
  `\&12\catcode `\#12\catcode `\^12\catcode `\_12\catcode `\%12\relax}%
\providecommand \@@startlink[1]{}%
\providecommand \@@endlink[0]{}%
\providecommand \url  [0]{\begingroup\@sanitize@url \@url }%
\providecommand \@url [1]{\endgroup\@href {#1}{\urlprefix }}%
\providecommand \urlprefix  [0]{URL }%
\providecommand \Eprint [0]{\href }%
\providecommand \doibase [0]{https://doi.org/}%
\providecommand \selectlanguage [0]{\@gobble}%
\providecommand \bibinfo  [0]{\@secondoftwo}%
\providecommand \bibfield  [0]{\@secondoftwo}%
\providecommand \translation [1]{[#1]}%
\providecommand \BibitemOpen [0]{}%
\providecommand \bibitemStop [0]{}%
\providecommand \bibitemNoStop [0]{.\EOS\space}%
\providecommand \EOS [0]{\spacefactor3000\relax}%
\providecommand \BibitemShut  [1]{\csname bibitem#1\endcsname}%
\let\auto@bib@innerbib\@empty
\bibitem [{\citenamefont {Beaurepaire}\ \emph {et~al.}(1996)\citenamefont
  {Beaurepaire}, \citenamefont {{J.-C. Merle}}, \citenamefont {Daunois},\ and\
  \citenamefont {{J.-Y. Bigot}}}]{Beaurepaire1996}%
  \BibitemOpen
  \bibfield  {author} {\bibinfo {author} {\bibfnamefont {E.}~\bibnamefont
  {Beaurepaire}}, \bibinfo {author} {\bibnamefont {{J.-C. Merle}}}, \bibinfo
  {author} {\bibfnamefont {A.}~\bibnamefont {Daunois}},\ and\ \bibinfo {author}
  {\bibnamefont {{J.-Y. Bigot}}},\ }\bibfield  {title} {\bibinfo {title}
  {Ultrafast spin dynamics in ferromagnetic {N}ickel},\ }\href
  {https://doi.org/10.1103/PhysRevLett.76.4250} {\bibfield  {journal} {\bibinfo
   {journal} {Phys. Rev. Lett.}\ }\textbf {\bibinfo {volume} {76}},\ \bibinfo
  {pages} {4250} (\bibinfo {year} {1996})}\BibitemShut {NoStop}%
\bibitem [{\citenamefont {{C.D. Stanciu}}\ \emph {et~al.}(2007)\citenamefont
  {{C.D. Stanciu}}, \citenamefont {Hansteen}, \citenamefont {{A.V. Kimel}},
  \citenamefont {Kirilyuk}, \citenamefont {Tsukamoto}, \citenamefont {Itoh},\
  and\ \citenamefont {{Th. Rasing}}}]{Stanciu2007}%
  \BibitemOpen
  \bibfield  {author} {\bibinfo {author} {\bibnamefont {{C.D. Stanciu}}},
  \bibinfo {author} {\bibfnamefont {F.}~\bibnamefont {Hansteen}}, \bibinfo
  {author} {\bibnamefont {{A.V. Kimel}}}, \bibinfo {author} {\bibfnamefont
  {A.}~\bibnamefont {Kirilyuk}}, \bibinfo {author} {\bibfnamefont
  {A.}~\bibnamefont {Tsukamoto}}, \bibinfo {author} {\bibfnamefont
  {A.}~\bibnamefont {Itoh}},\ and\ \bibinfo {author} {\bibnamefont {{Th.
  Rasing}}},\ }\bibfield  {title} {\bibinfo {title} {All-optical magnetic
  recording with circularly polarized light},\ }\href
  {https://doi.org/10.1103/PhysRevLett.99.047601} {\bibfield  {journal}
  {\bibinfo  {journal} {Phys. Rev. Lett.}\ }\textbf {\bibinfo {volume} {99}},\
  \bibinfo {pages} {047601} (\bibinfo {year} {2007})}\BibitemShut {NoStop}%
\bibitem [{\citenamefont {Malinowski}\ \emph {et~al.}(2008)\citenamefont
  {Malinowski}, \citenamefont {{F. Dalla Longa}}, \citenamefont {{J.H.H.
  Rietjens}}, \citenamefont {{P.V. Paluskar}}, \citenamefont {Huijink},
  \citenamefont {{H.J.M. Swagten}},\ and\ \citenamefont
  {Koopmans}}]{malinowski2008control}%
  \BibitemOpen
  \bibfield  {author} {\bibinfo {author} {\bibfnamefont {G.}~\bibnamefont
  {Malinowski}}, \bibinfo {author} {\bibnamefont {{F. Dalla Longa}}}, \bibinfo
  {author} {\bibnamefont {{J.H.H. Rietjens}}}, \bibinfo {author} {\bibnamefont
  {{P.V. Paluskar}}}, \bibinfo {author} {\bibfnamefont {R.}~\bibnamefont
  {Huijink}}, \bibinfo {author} {\bibnamefont {{H.J.M. Swagten}}},\ and\
  \bibinfo {author} {\bibfnamefont {B.}~\bibnamefont {Koopmans}},\ }\bibfield
  {title} {\bibinfo {title} {Control of speed and efficiency of ultrafast
  demagnetization by direct transfer of spin angular momentum},\ }\href
  {https://doi.org/10.1038/nphys1092} {\bibfield  {journal} {\bibinfo
  {journal} {Nat. Phys.}\ }\textbf {\bibinfo {volume} {4}},\ \bibinfo {pages}
  {855} (\bibinfo {year} {2008})}\BibitemShut {NoStop}%
\bibitem [{\citenamefont {Melnikov}\ \emph {et~al.}(2011)\citenamefont
  {Melnikov}, \citenamefont {Razdolski}, \citenamefont {{T.O. Wehling}},
  \citenamefont {{E.Th. Papaioannou}}, \citenamefont {Roddatis}, \citenamefont
  {Fumagalli}, \citenamefont {Aktsipetrov}, \citenamefont {{A.I.
  Lichtenstein}},\ and\ \citenamefont {Bovensiepen}}]{Melnikov2011}%
  \BibitemOpen
  \bibfield  {author} {\bibinfo {author} {\bibfnamefont {A.}~\bibnamefont
  {Melnikov}}, \bibinfo {author} {\bibfnamefont {I.}~\bibnamefont {Razdolski}},
  \bibinfo {author} {\bibnamefont {{T.O. Wehling}}}, \bibinfo {author}
  {\bibnamefont {{E.Th. Papaioannou}}}, \bibinfo {author} {\bibfnamefont
  {V.}~\bibnamefont {Roddatis}}, \bibinfo {author} {\bibfnamefont
  {P.}~\bibnamefont {Fumagalli}}, \bibinfo {author} {\bibfnamefont
  {O.}~\bibnamefont {Aktsipetrov}}, \bibinfo {author} {\bibnamefont {{A.I.
  Lichtenstein}}},\ and\ \bibinfo {author} {\bibfnamefont {U.}~\bibnamefont
  {Bovensiepen}},\ }\bibfield  {title} {\bibinfo {title} {Ultrafast transport
  of laser-excited spin-polarized carriers in {Au/Fe/MgO} (001)},\ }\href
  {https://doi.org/10.1103/PhysRevLett.107.076601} {\bibfield  {journal}
  {\bibinfo  {journal} {Phys. Rev. Lett.}\ }\textbf {\bibinfo {volume} {107}},\
  \bibinfo {pages} {076601} (\bibinfo {year} {2011})}\BibitemShut {NoStop}%
\bibitem [{\citenamefont {{G.-M. Choi}}\ \emph {et~al.}(2014)\citenamefont
  {{G.-M. Choi}}, \citenamefont {{B.-C. Min}}, \citenamefont {{K.-J. Lee}},\
  and\ \citenamefont {{D.G. Cahill}}}]{Choi2014}%
  \BibitemOpen
  \bibfield  {author} {\bibinfo {author} {\bibnamefont {{G.-M. Choi}}},
  \bibinfo {author} {\bibnamefont {{B.-C. Min}}}, \bibinfo {author}
  {\bibnamefont {{K.-J. Lee}}},\ and\ \bibinfo {author} {\bibnamefont {{D.G.
  Cahill}}},\ }\bibfield  {title} {\bibinfo {title} {Spin current generated by
  thermally driven ultrafast demagnetization},\ }\href
  {https://doi.org/10.1038/ncomms5334} {\bibfield  {journal} {\bibinfo
  {journal} {Nat. Commun.}\ }\textbf {\bibinfo {volume} {5}},\ \bibinfo {pages}
  {4334} (\bibinfo {year} {2014})}\BibitemShut {NoStop}%
\bibitem [{\citenamefont {{A.J. Schellekens}}\ \emph
  {et~al.}(2014)\citenamefont {{A.J. Schellekens}}, \citenamefont {{K.C.
  Kuiper}}, \citenamefont {{R.R.J.C. De Wit}},\ and\ \citenamefont
  {Koopmans}}]{Schellekens2014stt}%
  \BibitemOpen
  \bibfield  {author} {\bibinfo {author} {\bibnamefont {{A.J. Schellekens}}},
  \bibinfo {author} {\bibnamefont {{K.C. Kuiper}}}, \bibinfo {author}
  {\bibnamefont {{R.R.J.C. De Wit}}},\ and\ \bibinfo {author} {\bibfnamefont
  {B.}~\bibnamefont {Koopmans}},\ }\bibfield  {title} {\bibinfo {title}
  {Ultrafast spin-transfer torque driven by femtosecond pulsed-laser
  excitation},\ }\href {https://doi.org/10.1038/ncomms5333} {\bibfield
  {journal} {\bibinfo  {journal} {Nat. Commun.}\ }\textbf {\bibinfo {volume}
  {5}},\ \bibinfo {pages} {4333} (\bibinfo {year} {2014})}\BibitemShut
  {NoStop}%
\bibitem [{\citenamefont {Razdolski}\ \emph {et~al.}(2017)\citenamefont
  {Razdolski}, \citenamefont {Alekhin}, \citenamefont {Ilin}, \citenamefont
  {{J.P Meyburg}}, \citenamefont {Roddatis}, \citenamefont {Diesing},
  \citenamefont {Bovensiepen},\ and\ \citenamefont {Melnikov}}]{Razdolski2017}%
  \BibitemOpen
  \bibfield  {author} {\bibinfo {author} {\bibfnamefont {I.}~\bibnamefont
  {Razdolski}}, \bibinfo {author} {\bibfnamefont {A.}~\bibnamefont {Alekhin}},
  \bibinfo {author} {\bibfnamefont {N.}~\bibnamefont {Ilin}}, \bibinfo {author}
  {\bibnamefont {{J.P Meyburg}}}, \bibinfo {author} {\bibfnamefont
  {V.}~\bibnamefont {Roddatis}}, \bibinfo {author} {\bibfnamefont
  {D.}~\bibnamefont {Diesing}}, \bibinfo {author} {\bibfnamefont
  {U.}~\bibnamefont {Bovensiepen}},\ and\ \bibinfo {author} {\bibfnamefont
  {A.}~\bibnamefont {Melnikov}},\ }\bibfield  {title} {\bibinfo {title}
  {Nanoscale interface confinement of ultrafast spin transfer torque driving
  non-uniform spin dynamics},\ }\href {https://doi.org/10.1038/ncomms15007}
  {\bibfield  {journal} {\bibinfo  {journal} {Nat. Commun.}\ }\textbf {\bibinfo
  {volume} {8}},\ \bibinfo {pages} {15007} (\bibinfo {year}
  {2017})}\BibitemShut {NoStop}%
\bibitem [{\citenamefont {{M.L.M. Lalieu}}\ \emph {et~al.}(2017)\citenamefont
  {{M.L.M. Lalieu}}, \citenamefont {{P.L.J. Helgers}},\ and\ \citenamefont
  {Koopmans}}]{lalieu2017absorption}%
  \BibitemOpen
  \bibfield  {author} {\bibinfo {author} {\bibnamefont {{M.L.M. Lalieu}}},
  \bibinfo {author} {\bibnamefont {{P.L.J. Helgers}}},\ and\ \bibinfo {author}
  {\bibfnamefont {B.}~\bibnamefont {Koopmans}},\ }\bibfield  {title} {\bibinfo
  {title} {Absorption and generation of femtosecond laser-pulse excited spin
  currents in noncollinear magnetic bilayers},\ }\href
  {https://doi.org/10.1103/PhysRevB.96.014417} {\bibfield  {journal} {\bibinfo
  {journal} {Phys. Rev. B}\ }\textbf {\bibinfo {volume} {96}},\ \bibinfo
  {pages} {014417} (\bibinfo {year} {2017})}\BibitemShut {NoStop}%
\bibitem [{\citenamefont {{G.P. Zhang}}\ and\ \citenamefont
  {H{\"u}bner}(2000)}]{Zhang2000}%
  \BibitemOpen
  \bibfield  {author} {\bibinfo {author} {\bibnamefont {{G.P. Zhang}}}\ and\
  \bibinfo {author} {\bibfnamefont {W.}~\bibnamefont {H{\"u}bner}},\ }\bibfield
   {title} {\bibinfo {title} {Laser-induced ultrafast demagnetization in
  ferromagnetic metals},\ }\href {https://doi.org/10.1103/PhysRevLett.85.3025}
  {\bibfield  {journal} {\bibinfo  {journal} {Phys. Rev. Lett.}\ }\textbf
  {\bibinfo {volume} {85}},\ \bibinfo {pages} {3025} (\bibinfo {year}
  {2000})}\BibitemShut {NoStop}%
\bibitem [{\citenamefont {{J.-Y. Bigot}}\ \emph {et~al.}(2009)\citenamefont
  {{J.-Y. Bigot}}, \citenamefont {Vomir},\ and\ \citenamefont
  {Beaurepaire}}]{Bigot2009}%
  \BibitemOpen
  \bibfield  {author} {\bibinfo {author} {\bibnamefont {{J.-Y. Bigot}}},
  \bibinfo {author} {\bibfnamefont {M.}~\bibnamefont {Vomir}},\ and\ \bibinfo
  {author} {\bibfnamefont {E.}~\bibnamefont {Beaurepaire}},\ }\bibfield
  {title} {\bibinfo {title} {Coherent ultrafast magnetism induced by
  femtosecond laser pulses},\ }\href {https://doi.org/10.1038/nphys1285}
  {\bibfield  {journal} {\bibinfo  {journal} {Nat. Phys.}\ }\textbf {\bibinfo
  {volume} {5}},\ \bibinfo {pages} {515} (\bibinfo {year} {2009})}\BibitemShut
  {NoStop}%
\bibitem [{\citenamefont {Battiato}\ \emph {et~al.}(2010)\citenamefont
  {Battiato}, \citenamefont {Carva},\ and\ \citenamefont
  {Oppeneer}}]{Battiato2010}%
  \BibitemOpen
  \bibfield  {author} {\bibinfo {author} {\bibfnamefont {M.}~\bibnamefont
  {Battiato}}, \bibinfo {author} {\bibfnamefont {K.}~\bibnamefont {Carva}},\
  and\ \bibinfo {author} {\bibfnamefont {P.~M.}\ \bibnamefont {Oppeneer}},\
  }\bibfield  {title} {\bibinfo {title} {Superdiffusive spin transport as a
  mechanism of ultrafast demagnetization},\ }\href
  {https://doi.org/10.1103/PhysRevLett.105.027203} {\bibfield  {journal}
  {\bibinfo  {journal} {Phys. Rev. Lett.}\ }\textbf {\bibinfo {volume} {105}},\
  \bibinfo {pages} {027203} (\bibinfo {year} {2010})}\BibitemShut {NoStop}%
\bibitem [{\citenamefont {Koopmans}\ \emph {et~al.}(2005)\citenamefont
  {Koopmans}, \citenamefont {{J.J.M. Ruigrok}}, \citenamefont {{Dalla Longa}},\
  and\ \citenamefont {{W.J.M. de Jonge}}}]{Koopmans2005}%
  \BibitemOpen
  \bibfield  {author} {\bibinfo {author} {\bibfnamefont {B.}~\bibnamefont
  {Koopmans}}, \bibinfo {author} {\bibnamefont {{J.J.M. Ruigrok}}}, \bibinfo
  {author} {\bibfnamefont {F.}~\bibnamefont {{Dalla Longa}}},\ and\ \bibinfo
  {author} {\bibnamefont {{W.J.M. de Jonge}}},\ }\bibfield  {title}
  {{\selectlanguage {English}\bibinfo {title} {Unifying ultrafast magnetization
  dynamics}},\ }\href {https://doi.org/10.1103/PhysRevLett.95.267207}
  {\bibfield  {journal} {\bibinfo  {journal} {Phys. Rev. Lett.}\ }\textbf
  {\bibinfo {volume} {95}},\ \bibinfo {pages} {267207} (\bibinfo {year}
  {2005})}\BibitemShut {NoStop}%
\bibitem [{\citenamefont {Kazantseva}\ \emph {et~al.}(2007)\citenamefont
  {Kazantseva}, \citenamefont {Nowak}, \citenamefont {{R.W. Chantrell}},
  \citenamefont {Hohlfeld},\ and\ \citenamefont {Rebei}}]{Kazantseva2007}%
  \BibitemOpen
  \bibfield  {author} {\bibinfo {author} {\bibfnamefont {N.}~\bibnamefont
  {Kazantseva}}, \bibinfo {author} {\bibfnamefont {U.}~\bibnamefont {Nowak}},
  \bibinfo {author} {\bibnamefont {{R.W. Chantrell}}}, \bibinfo {author}
  {\bibfnamefont {J.}~\bibnamefont {Hohlfeld}},\ and\ \bibinfo {author}
  {\bibfnamefont {A.}~\bibnamefont {Rebei}},\ }\bibfield  {title} {\bibinfo
  {title} {Slow recovery of the magnetisation after a sub-picosecond heat
  pulse},\ }\href {https://doi.org/10.1209/0295-5075/81/27004} {\bibfield
  {journal} {\bibinfo  {journal} {Europhys. Lett.}\ }\textbf {\bibinfo {volume}
  {81}},\ \bibinfo {pages} {27004} (\bibinfo {year} {2007})}\BibitemShut
  {NoStop}%
\bibitem [{\citenamefont {Krau{\ss}}\ \emph {et~al.}(2009)\citenamefont
  {Krau{\ss}}, \citenamefont {Roth}, \citenamefont {Alebrand}, \citenamefont
  {Steil}, \citenamefont {Cinchetti}, \citenamefont {Aeschlimann},\ and\
  \citenamefont {{H.C. Schneider}}}]{Krauss2009}%
  \BibitemOpen
  \bibfield  {author} {\bibinfo {author} {\bibfnamefont {M.}~\bibnamefont
  {Krau{\ss}}}, \bibinfo {author} {\bibfnamefont {T.}~\bibnamefont {Roth}},
  \bibinfo {author} {\bibfnamefont {S.}~\bibnamefont {Alebrand}}, \bibinfo
  {author} {\bibfnamefont {D.}~\bibnamefont {Steil}}, \bibinfo {author}
  {\bibfnamefont {M.}~\bibnamefont {Cinchetti}}, \bibinfo {author}
  {\bibfnamefont {M.}~\bibnamefont {Aeschlimann}},\ and\ \bibinfo {author}
  {\bibnamefont {{H.C. Schneider}}},\ }\bibfield  {title} {\bibinfo {title}
  {Ultrafast demagnetization of ferromagnetic transition metals: The role of
  the {C}oulomb interaction},\ }\href
  {https://doi.org/10.1103/PhysRevB.80.180407} {\bibfield  {journal} {\bibinfo
  {journal} {Phys. Rev. B}\ }\textbf {\bibinfo {volume} {80}},\ \bibinfo
  {pages} {180407(R)} (\bibinfo {year} {2009})}\BibitemShut {NoStop}%
\bibitem [{\citenamefont {Koopmans}\ \emph {et~al.}(2010)\citenamefont
  {Koopmans}, \citenamefont {Malinowski}, \citenamefont {{Dalla Longa}},
  \citenamefont {Steiauf}, \citenamefont {F{\"a}hnle}, \citenamefont {Roth},
  \citenamefont {Cinchetti},\ and\ \citenamefont {Aeschlimann}}]{Koopmans2010}%
  \BibitemOpen
  \bibfield  {author} {\bibinfo {author} {\bibfnamefont {B.}~\bibnamefont
  {Koopmans}}, \bibinfo {author} {\bibfnamefont {G.}~\bibnamefont
  {Malinowski}}, \bibinfo {author} {\bibfnamefont {F.}~\bibnamefont {{Dalla
  Longa}}}, \bibinfo {author} {\bibfnamefont {D.}~\bibnamefont {Steiauf}},
  \bibinfo {author} {\bibfnamefont {M.}~\bibnamefont {F{\"a}hnle}}, \bibinfo
  {author} {\bibfnamefont {T.}~\bibnamefont {Roth}}, \bibinfo {author}
  {\bibfnamefont {M.}~\bibnamefont {Cinchetti}},\ and\ \bibinfo {author}
  {\bibfnamefont {M.}~\bibnamefont {Aeschlimann}},\ }\bibfield  {title}
  {{\selectlanguage {English}\bibinfo {title} {Explaining the paradoxical
  diversity of ultrafast laser-induced demagnetization}},\ }\href
  {https://doi.org/10.1038/NMAT2593} {\bibfield  {journal} {\bibinfo  {journal}
  {Nat. Mat.}\ }\textbf {\bibinfo {volume} {9}},\ \bibinfo {pages} {259}
  (\bibinfo {year} {2010})}\BibitemShut {NoStop}%
\bibitem [{\citenamefont {Manchon}\ \emph {et~al.}(2012)\citenamefont
  {Manchon}, \citenamefont {Li}, \citenamefont {Xu},\ and\ \citenamefont
  {Zhang}}]{Manchon2012}%
  \BibitemOpen
  \bibfield  {author} {\bibinfo {author} {\bibfnamefont {A.}~\bibnamefont
  {Manchon}}, \bibinfo {author} {\bibfnamefont {Q.}~\bibnamefont {Li}},
  \bibinfo {author} {\bibfnamefont {L.}~\bibnamefont {Xu}},\ and\ \bibinfo
  {author} {\bibfnamefont {S.}~\bibnamefont {Zhang}},\ }\bibfield  {title}
  {\bibinfo {title} {Theory of laser-induced demagnetization at high
  temperatures},\ }\href {https://doi.org/10.1103/PhysRevB.85.064408}
  {\bibfield  {journal} {\bibinfo  {journal} {Phys. Rev. B}\ }\textbf {\bibinfo
  {volume} {85}},\ \bibinfo {pages} {064408} (\bibinfo {year}
  {2012})}\BibitemShut {NoStop}%
\bibitem [{\citenamefont {{B.Y. Mueller}}\ and\ \citenamefont
  {Rethfeld}(2014)}]{Mueller2014}%
  \BibitemOpen
  \bibfield  {author} {\bibinfo {author} {\bibnamefont {{B.Y. Mueller}}}\ and\
  \bibinfo {author} {\bibfnamefont {B.}~\bibnamefont {Rethfeld}},\ }\bibfield
  {title} {\bibinfo {title} {Thermodynamic $\mu${T} model of ultrafast
  magnetization dynamics},\ }\href {https://doi.org/10.1103/PhysRevB.90.144420}
  {\bibfield  {journal} {\bibinfo  {journal} {Phys. Rev. B}\ }\textbf {\bibinfo
  {volume} {90}},\ \bibinfo {pages} {144420} (\bibinfo {year}
  {2014})}\BibitemShut {NoStop}%
\bibitem [{\citenamefont {Nieves}\ \emph {et~al.}(2014)\citenamefont {Nieves},
  \citenamefont {Serantes}, \citenamefont {Atxitia},\ and\ \citenamefont
  {Chubykalo-Fesenko}}]{Nieves2014}%
  \BibitemOpen
  \bibfield  {author} {\bibinfo {author} {\bibfnamefont {P.}~\bibnamefont
  {Nieves}}, \bibinfo {author} {\bibfnamefont {D.}~\bibnamefont {Serantes}},
  \bibinfo {author} {\bibfnamefont {U.}~\bibnamefont {Atxitia}},\ and\ \bibinfo
  {author} {\bibfnamefont {O.}~\bibnamefont {Chubykalo-Fesenko}},\ }\bibfield
  {title} {\bibinfo {title} {Quantum {L}andau-{L}ifshitz-{B}loch equation and
  its comparison with the classical case},\ }\href
  {https://doi.org/10.1103/PhysRevB.90.104428} {\bibfield  {journal} {\bibinfo
  {journal} {Phys. Rev. B}\ }\textbf {\bibinfo {volume} {90}},\ \bibinfo
  {pages} {104428} (\bibinfo {year} {2014})}\BibitemShut {NoStop}%
\bibitem [{\citenamefont {Tveten}\ \emph {et~al.}(2015)\citenamefont {Tveten},
  \citenamefont {Brataas},\ and\ \citenamefont {Tserkovnyak}}]{Tveten2015}%
  \BibitemOpen
  \bibfield  {author} {\bibinfo {author} {\bibfnamefont {E.~G.}\ \bibnamefont
  {Tveten}}, \bibinfo {author} {\bibfnamefont {A.}~\bibnamefont {Brataas}},\
  and\ \bibinfo {author} {\bibfnamefont {Y.}~\bibnamefont {Tserkovnyak}},\
  }\bibfield  {title} {\bibinfo {title} {Electron-magnon scattering in magnetic
  heterostructures far out of equilibrium},\ }\href
  {https://doi.org/10.1103/PhysRevB.92.180412} {\bibfield  {journal} {\bibinfo
  {journal} {Phys. Rev. B}\ }\textbf {\bibinfo {volume} {92}},\ \bibinfo
  {pages} {180412(R)} (\bibinfo {year} {2015})}\BibitemShut {NoStop}%
\bibitem [{\citenamefont {Krieger}\ \emph {et~al.}(2015)\citenamefont
  {Krieger}, \citenamefont {{J.K. Dewhurst}}, \citenamefont {Elliott},
  \citenamefont {Sharma},\ and\ \citenamefont {{E.K.U. Gross}}}]{Krieger2015}%
  \BibitemOpen
  \bibfield  {author} {\bibinfo {author} {\bibfnamefont {K.}~\bibnamefont
  {Krieger}}, \bibinfo {author} {\bibnamefont {{J.K. Dewhurst}}}, \bibinfo
  {author} {\bibfnamefont {P.}~\bibnamefont {Elliott}}, \bibinfo {author}
  {\bibfnamefont {S.}~\bibnamefont {Sharma}},\ and\ \bibinfo {author}
  {\bibnamefont {{E.K.U. Gross}}},\ }\bibfield  {title} {\bibinfo {title}
  {Laser-induced demagnetization at ultrashort time scales: Predictions of
  {TDDFT}},\ }\href {https://doi.org/10.1021/acs.jctc.5b00621} {\bibfield
  {journal} {\bibinfo  {journal} {Journal of chemical theory and computation}\
  }\textbf {\bibinfo {volume} {11}},\ \bibinfo {pages} {4870} (\bibinfo {year}
  {2015})}\BibitemShut {NoStop}%
\bibitem [{\citenamefont {T{\"o}ws}\ and\ \citenamefont {{G.M.
  Pastor}}(2015)}]{Tows2015}%
  \BibitemOpen
  \bibfield  {author} {\bibinfo {author} {\bibfnamefont {W.}~\bibnamefont
  {T{\"o}ws}}\ and\ \bibinfo {author} {\bibnamefont {{G.M. Pastor}}},\
  }\bibfield  {title} {\bibinfo {title} {Many-body theory of ultrafast
  demagnetization and angular momentum transfer in ferromagnetic transition
  metals},\ }\href {https://doi.org/10.1103/PhysRevLett.115.217204} {\bibfield
  {journal} {\bibinfo  {journal} {Phys. Rev. Lett.}\ }\textbf {\bibinfo
  {volume} {115}},\ \bibinfo {pages} {217204} (\bibinfo {year}
  {2015})}\BibitemShut {NoStop}%
\bibitem [{\citenamefont {Cywi{\'n}ski}\ and\ \citenamefont {{L.J.
  Sham}}(2007)}]{Cywinski2007}%
  \BibitemOpen
  \bibfield  {author} {\bibinfo {author} {\bibfnamefont {{\L}.}~\bibnamefont
  {Cywi{\'n}ski}}\ and\ \bibinfo {author} {\bibnamefont {{L.J. Sham}}},\
  }\bibfield  {title} {\bibinfo {title} {Ultrafast demagnetization in the sp-d
  model: A theoretical study},\ }\href
  {https://doi.org/10.1103/PhysRevB.76.045205} {\bibfield  {journal} {\bibinfo
  {journal} {Phys. Rev. B}\ }\textbf {\bibinfo {volume} {76}},\ \bibinfo
  {pages} {045205} (\bibinfo {year} {2007})}\BibitemShut {NoStop}%
\bibitem [{\citenamefont {{V.N. Gridnev}}(2016)}]{Gridnev2016}%
  \BibitemOpen
  \bibfield  {author} {\bibinfo {author} {\bibnamefont {{V.N. Gridnev}}},\
  }\bibfield  {title} {\bibinfo {title} {Ultrafast heating-induced
  magnetization switching in ferrimagnets},\ }\href
  {https://doi.org/10.1088/0953-8984/28/47/476007} {\bibfield  {journal}
  {\bibinfo  {journal} {Journal of Physics: Condensed Matter}\ }\textbf
  {\bibinfo {volume} {28}},\ \bibinfo {pages} {476007} (\bibinfo {year}
  {2016})}\BibitemShut {NoStop}%
\bibitem [{\citenamefont {Battiato}\ \emph {et~al.}(2012)\citenamefont
  {Battiato}, \citenamefont {Carva},\ and\ \citenamefont {{P.M.
  Oppeneer}}}]{Battiato2012}%
  \BibitemOpen
  \bibfield  {author} {\bibinfo {author} {\bibfnamefont {M.}~\bibnamefont
  {Battiato}}, \bibinfo {author} {\bibfnamefont {K.}~\bibnamefont {Carva}},\
  and\ \bibinfo {author} {\bibnamefont {{P.M. Oppeneer}}},\ }\bibfield  {title}
  {\bibinfo {title} {Theory of laser-induced ultrafast superdiffusive spin
  transport in layered heterostructures},\ }\href
  {https://doi.org/10.1103/PhysRevB.86.024404} {\bibfield  {journal} {\bibinfo
  {journal} {Phys. Rev. B}\ }\textbf {\bibinfo {volume} {86}},\ \bibinfo
  {pages} {024404} (\bibinfo {year} {2012})}\BibitemShut {NoStop}%
\bibitem [{\citenamefont {Choi}\ \emph {et~al.}(2015)\citenamefont {Choi},
  \citenamefont {Moon}, \citenamefont {Min}, \citenamefont {Lee},\ and\
  \citenamefont {{D.G. Cahill}}}]{Choi2015}%
  \BibitemOpen
  \bibfield  {author} {\bibinfo {author} {\bibfnamefont {G.-M.}\ \bibnamefont
  {Choi}}, \bibinfo {author} {\bibfnamefont {C.-H.}\ \bibnamefont {Moon}},
  \bibinfo {author} {\bibfnamefont {B.-C.}\ \bibnamefont {Min}}, \bibinfo
  {author} {\bibfnamefont {K.-J.}\ \bibnamefont {Lee}},\ and\ \bibinfo {author}
  {\bibnamefont {{D.G. Cahill}}},\ }\bibfield  {title} {\bibinfo {title}
  {Thermal spin-transfer torque driven by the spin-dependent {S}eebeck effect
  in metallic spin-valves},\ }\href {https://doi.org/10.1038/nphys3355}
  {\bibfield  {journal} {\bibinfo  {journal} {Nat. Phys.}\ }\textbf {\bibinfo
  {volume} {11}},\ \bibinfo {pages} {576} (\bibinfo {year} {2015})}\BibitemShut
  {NoStop}%
\bibitem [{\citenamefont {Alekhin}\ \emph {et~al.}(2017)\citenamefont
  {Alekhin}, \citenamefont {Razdolski}, \citenamefont {Ilin}, \citenamefont
  {{J.P. Meyburg}}, \citenamefont {Diesing}, \citenamefont {Roddatis},
  \citenamefont {Rungger}, \citenamefont {Stamenova}, \citenamefont {Sanvito},
  \citenamefont {Bovensiepen},\ and\ \citenamefont {Melnikov}}]{Alekhin2017}%
  \BibitemOpen
  \bibfield  {author} {\bibinfo {author} {\bibfnamefont {A.}~\bibnamefont
  {Alekhin}}, \bibinfo {author} {\bibfnamefont {I.}~\bibnamefont {Razdolski}},
  \bibinfo {author} {\bibfnamefont {N.}~\bibnamefont {Ilin}}, \bibinfo {author}
  {\bibnamefont {{J.P. Meyburg}}}, \bibinfo {author} {\bibfnamefont
  {D.}~\bibnamefont {Diesing}}, \bibinfo {author} {\bibfnamefont
  {V.}~\bibnamefont {Roddatis}}, \bibinfo {author} {\bibfnamefont
  {I.}~\bibnamefont {Rungger}}, \bibinfo {author} {\bibfnamefont
  {M.}~\bibnamefont {Stamenova}}, \bibinfo {author} {\bibfnamefont
  {S.}~\bibnamefont {Sanvito}}, \bibinfo {author} {\bibfnamefont
  {U.}~\bibnamefont {Bovensiepen}},\ and\ \bibinfo {author} {\bibfnamefont
  {A.}~\bibnamefont {Melnikov}},\ }\bibfield  {title} {\bibinfo {title}
  {Femtosecond spin current pulses generated by the nonthermal spin-dependent
  {S}eebeck effect and interacting with ferromagnets in spin valves},\ }\href
  {https://doi.org/10.1103/PhysRevLett.119.017202} {\bibfield  {journal}
  {\bibinfo  {journal} {Phys. Rev. Lett.}\ }\textbf {\bibinfo {volume} {119}},\
  \bibinfo {pages} {017202} (\bibinfo {year} {2017})}\BibitemShut {NoStop}%
\bibitem [{\citenamefont {Shin}\ \emph {et~al.}(2018)\citenamefont {Shin},
  \citenamefont {Min}, \citenamefont {Ju},\ and\ \citenamefont
  {Choi}}]{Shin2018}%
  \BibitemOpen
  \bibfield  {author} {\bibinfo {author} {\bibfnamefont {I.-H.}\ \bibnamefont
  {Shin}}, \bibinfo {author} {\bibfnamefont {B.-C.}\ \bibnamefont {Min}},
  \bibinfo {author} {\bibfnamefont {B.-K.}\ \bibnamefont {Ju}},\ and\ \bibinfo
  {author} {\bibfnamefont {G.-M.}\ \bibnamefont {Choi}},\ }\bibfield  {title}
  {\bibinfo {title} {Ultrafast spin current generated by electron--magnon
  scattering in bulk of ferromagnets},\ }\href
  {https://doi.org/10.7567/JJAP.57.090307} {\bibfield  {journal} {\bibinfo
  {journal} {Japanese Journal of Applied Physics}\ }\textbf {\bibinfo {volume}
  {57}},\ \bibinfo {pages} {090307} (\bibinfo {year} {2018})}\BibitemShut
  {NoStop}%
\bibitem [{\citenamefont {Kimling}\ and\ \citenamefont {{D.G.
  Cahill}}(2017)}]{Kimling2017}%
  \BibitemOpen
  \bibfield  {author} {\bibinfo {author} {\bibfnamefont {J.}~\bibnamefont
  {Kimling}}\ and\ \bibinfo {author} {\bibnamefont {{D.G. Cahill}}},\
  }\bibfield  {title} {\bibinfo {title} {Spin diffusion induced by pulsed-laser
  heating and the role of spin heat accumulation},\ }\href
  {https://doi.org/10.1103/PhysRevB.95.014402} {\bibfield  {journal} {\bibinfo
  {journal} {Phys. Rev. B}\ }\textbf {\bibinfo {volume} {95}},\ \bibinfo
  {pages} {014402} (\bibinfo {year} {2017})}\BibitemShut {NoStop}%
\bibitem [{\citenamefont {{R.C. Iotti}}\ \emph {et~al.}(2005)\citenamefont
  {{R.C. Iotti}}, \citenamefont {Ciancio},\ and\ \citenamefont
  {Rossi}}]{Iotti2005}%
  \BibitemOpen
  \bibfield  {author} {\bibinfo {author} {\bibnamefont {{R.C. Iotti}}},
  \bibinfo {author} {\bibfnamefont {E.}~\bibnamefont {Ciancio}},\ and\ \bibinfo
  {author} {\bibfnamefont {F.}~\bibnamefont {Rossi}},\ }\bibfield  {title}
  {\bibinfo {title} {Quantum transport theory for semiconductor nanostructures:
  A density-matrix formulation},\ }\href
  {https://doi.org/10.1103/PhysRevB.72.125347} {\bibfield  {journal} {\bibinfo
  {journal} {Phys. Rev. B}\ }\textbf {\bibinfo {volume} {72}},\ \bibinfo
  {pages} {125347} (\bibinfo {year} {2005})}\BibitemShut {NoStop}%
\bibitem [{\citenamefont {{S.I. Anisimov}}\ \emph {et~al.}(1974)\citenamefont
  {{S.I. Anisimov}}, \citenamefont {{B.L. Kapeliovich}},\ and\ \citenamefont
  {{T.L. Perelman}}}]{Anisimov1974}%
  \BibitemOpen
  \bibfield  {author} {\bibinfo {author} {\bibnamefont {{S.I. Anisimov}}},
  \bibinfo {author} {\bibnamefont {{B.L. Kapeliovich}}},\ and\ \bibinfo
  {author} {\bibnamefont {{T.L. Perelman}}},\ }\bibfield  {title} {\bibinfo
  {title} {Electron emission from metal surfaces exposed to ultrashort laser
  pulses},\ }\href@noop {} {\bibfield  {journal} {\bibinfo  {journal} {Zh.
  Eksp. Teor. Fiz}\ }\textbf {\bibinfo {volume} {66}},\ \bibinfo {pages} {375}
  (\bibinfo {year} {1974})}\BibitemShut {NoStop}%
\bibitem [{Foo()}]{Footnote}%
  \BibitemOpen
  \href@noop {} {}\bibinfo {note} {The time scale $\tau_M$ corresponds to the
  prefactor $\tau_M^{-1}= R T_p/T_C$ from the standard M3TM
  \cite{Koopmans2010}, and is taken as a constant.}\BibitemShut {Stop}%
\bibitem [{\citenamefont {Rudolf}\ \emph {et~al.}(2012)\citenamefont {Rudolf},
  \citenamefont {{C. La-O-Vorakiat}}, \citenamefont {Battiato}, \citenamefont
  {Adam}, \citenamefont {{J.M. Shaw}}, \citenamefont {Turgut}, \citenamefont
  {Maldonado}, \citenamefont {Mathias}, \citenamefont {Grychtol}, \citenamefont
  {{H.T. Nembach}}, \citenamefont {{T.J. Silva}}, \citenamefont {{M.
  Aeschlimann}}, \citenamefont {{H.C. Kapteyn}}, \citenamefont {{M.M.
  Murnane}}, \citenamefont {{C.M. Schneider}},\ and\ \citenamefont {{P.M.
  Oppeneer}}}]{Rudolf2012}%
  \BibitemOpen
  \bibfield  {author} {\bibinfo {author} {\bibfnamefont {D.}~\bibnamefont
  {Rudolf}}, \bibinfo {author} {\bibnamefont {{C. La-O-Vorakiat}}}, \bibinfo
  {author} {\bibfnamefont {M.}~\bibnamefont {Battiato}}, \bibinfo {author}
  {\bibfnamefont {R.}~\bibnamefont {Adam}}, \bibinfo {author} {\bibnamefont
  {{J.M. Shaw}}}, \bibinfo {author} {\bibfnamefont {E.}~\bibnamefont {Turgut}},
  \bibinfo {author} {\bibfnamefont {P.}~\bibnamefont {Maldonado}}, \bibinfo
  {author} {\bibfnamefont {S.}~\bibnamefont {Mathias}}, \bibinfo {author}
  {\bibfnamefont {P.}~\bibnamefont {Grychtol}}, \bibinfo {author} {\bibnamefont
  {{H.T. Nembach}}}, \bibinfo {author} {\bibnamefont {{T.J. Silva}}}, \bibinfo
  {author} {\bibnamefont {{M. Aeschlimann}}}, \bibinfo {author} {\bibnamefont
  {{H.C. Kapteyn}}}, \bibinfo {author} {\bibnamefont {{M.M. Murnane}}},
  \bibinfo {author} {\bibnamefont {{C.M. Schneider}}},\ and\ \bibinfo {author}
  {\bibnamefont {{P.M. Oppeneer}}},\ }\bibfield  {title} {\bibinfo {title}
  {Ultrafast magnetization enhancement in metallic multilayers driven by
  superdiffusive spin current},\ }\href {https://doi.org/10.1038/ncomms2029}
  {\bibfield  {journal} {\bibinfo  {journal} {Nat. Commun.}\ }\textbf {\bibinfo
  {volume} {3}},\ \bibinfo {pages} {1} (\bibinfo {year} {2012})}\BibitemShut
  {NoStop}%
\bibitem [{\citenamefont {Slachter}\ \emph {et~al.}(2010)\citenamefont
  {Slachter}, \citenamefont {{F.L. Bakker}}, \citenamefont {{J.-P. Adam}},\
  and\ \citenamefont {{B.J. van Wees}}}]{Slachter2010}%
  \BibitemOpen
  \bibfield  {author} {\bibinfo {author} {\bibfnamefont {A.}~\bibnamefont
  {Slachter}}, \bibinfo {author} {\bibnamefont {{F.L. Bakker}}}, \bibinfo
  {author} {\bibnamefont {{J.-P. Adam}}},\ and\ \bibinfo {author} {\bibnamefont
  {{B.J. van Wees}}},\ }\bibfield  {title} {\bibinfo {title} {Thermally driven
  spin injection from a ferromagnet into a non-magnetic metal},\ }\href
  {https://doi.org/10.1038/nphys1767} {\bibfield  {journal} {\bibinfo
  {journal} {Nat. Phys.}\ }\textbf {\bibinfo {volume} {6}},\ \bibinfo {pages}
  {879} (\bibinfo {year} {2010})}\BibitemShut {NoStop}%
\bibitem [{\citenamefont {Beens}\ \emph {et~al.}(2018)\citenamefont {Beens},
  \citenamefont {{J.P. Heremans}}, \citenamefont {Tserkovnyak},\ and\
  \citenamefont {{R.A. Duine}}}]{Beens2018}%
  \BibitemOpen
  \bibfield  {author} {\bibinfo {author} {\bibfnamefont {M.}~\bibnamefont
  {Beens}}, \bibinfo {author} {\bibnamefont {{J.P. Heremans}}}, \bibinfo
  {author} {\bibfnamefont {Y.}~\bibnamefont {Tserkovnyak}},\ and\ \bibinfo
  {author} {\bibnamefont {{R.A. Duine}}},\ }\bibfield  {title} {\bibinfo
  {title} {Magnons versus electrons in thermal spin transport through metallic
  interfaces},\ }\href {https://doi.org/10.1088/1361-6463/aad520} {\bibfield
  {journal} {\bibinfo  {journal} {Journal of Physics D: Applied Physics}\
  }\textbf {\bibinfo {volume} {51}},\ \bibinfo {pages} {394002} (\bibinfo
  {year} {2018})}\BibitemShut {NoStop}%
\bibitem [{\citenamefont {{L.J. Cornelissen}}\ \emph
  {et~al.}(2016)\citenamefont {{L.J. Cornelissen}}, \citenamefont {{K.J.H.
  Peters}}, \citenamefont {{G.E.W. Bauer}}, \citenamefont {{R.A. Duine}},\ and\
  \citenamefont {{B.J. van Wees}}}]{cornelissen2016magnon}%
  \BibitemOpen
  \bibfield  {author} {\bibinfo {author} {\bibnamefont {{L.J. Cornelissen}}},
  \bibinfo {author} {\bibnamefont {{K.J.H. Peters}}}, \bibinfo {author}
  {\bibnamefont {{G.E.W. Bauer}}}, \bibinfo {author} {\bibnamefont {{R.A.
  Duine}}},\ and\ \bibinfo {author} {\bibnamefont {{B.J. van Wees}}},\
  }\bibfield  {title} {\bibinfo {title} {Magnon spin transport driven by the
  magnon chemical potential in a magnetic insulator},\ }\href
  {https://doi.org/10.1103/PhysRevB.94.014412} {\bibfield  {journal} {\bibinfo
  {journal} {Phys. Rev. B}\ }\textbf {\bibinfo {volume} {94}},\ \bibinfo
  {pages} {014412} (\bibinfo {year} {2016})}\BibitemShut {NoStop}%
\bibitem [{\citenamefont {Uchida}\ \emph {et~al.}(2010)\citenamefont {Uchida},
  \citenamefont {Xiao}, \citenamefont {Adachi}, \citenamefont {Ohe},
  \citenamefont {Takahashi}, \citenamefont {Ieda}, \citenamefont {Ota},
  \citenamefont {Kajiwara}, \citenamefont {Umezawa}, \citenamefont {Kawai},
  \citenamefont {{G.E.W. Bauer}}, \citenamefont {{S. Maekawa}},\ and\
  \citenamefont {{E. Saitoh}}}]{Uchida2010}%
  \BibitemOpen
  \bibfield  {author} {\bibinfo {author} {\bibfnamefont {K.}~\bibnamefont
  {Uchida}}, \bibinfo {author} {\bibfnamefont {J.}~\bibnamefont {Xiao}},
  \bibinfo {author} {\bibfnamefont {H.}~\bibnamefont {Adachi}}, \bibinfo
  {author} {\bibfnamefont {J.}~\bibnamefont {Ohe}}, \bibinfo {author}
  {\bibfnamefont {S.}~\bibnamefont {Takahashi}}, \bibinfo {author}
  {\bibfnamefont {J.}~\bibnamefont {Ieda}}, \bibinfo {author} {\bibfnamefont
  {T.}~\bibnamefont {Ota}}, \bibinfo {author} {\bibfnamefont {Y.}~\bibnamefont
  {Kajiwara}}, \bibinfo {author} {\bibfnamefont {H.}~\bibnamefont {Umezawa}},
  \bibinfo {author} {\bibfnamefont {H.}~\bibnamefont {Kawai}}, \bibinfo
  {author} {\bibnamefont {{G.E.W. Bauer}}}, \bibinfo {author} {\bibnamefont
  {{S. Maekawa}}},\ and\ \bibinfo {author} {\bibnamefont {{E. Saitoh}}},\
  }\bibfield  {title} {\bibinfo {title} {Spin {S}eebeck insulator},\ }\href
  {https://doi.org/10.1038/nmat2856} {\bibfield  {journal} {\bibinfo  {journal}
  {Nat. Mat.}\ }\textbf {\bibinfo {volume} {9}},\ \bibinfo {pages} {894}
  (\bibinfo {year} {2010})}\BibitemShut {NoStop}%
\bibitem [{\citenamefont {Xiao}\ \emph {et~al.}(2010)\citenamefont {Xiao},
  \citenamefont {{G.E.W. Bauer}}, \citenamefont {{K.C. Uchida}}, \citenamefont
  {Saitoh},\ and\ \citenamefont {Maekawa}}]{Xiao2010}%
  \BibitemOpen
  \bibfield  {author} {\bibinfo {author} {\bibfnamefont {J.}~\bibnamefont
  {Xiao}}, \bibinfo {author} {\bibnamefont {{G.E.W. Bauer}}}, \bibinfo {author}
  {\bibnamefont {{K.C. Uchida}}}, \bibinfo {author} {\bibfnamefont
  {E.}~\bibnamefont {Saitoh}},\ and\ \bibinfo {author} {\bibfnamefont
  {S.}~\bibnamefont {Maekawa}},\ }\bibfield  {title} {\bibinfo {title} {Theory
  of magnon-driven spin {S}eebeck effect},\ }\href
  {https://doi.org/10.1103/PhysRevB.81.214418} {\bibfield  {journal} {\bibinfo
  {journal} {Phys. Rev. B}\ }\textbf {\bibinfo {volume} {81}},\ \bibinfo
  {pages} {214418} (\bibinfo {year} {2010})}\BibitemShut {NoStop}%
\end{thebibliography}

\providecommand{\noopsort}[1]{}\providecommand{\singleletter}[1]{#1}%

\end{document}